THIS IS A LaTeX FILE.  RUN LaTeX TWICE TO GET CROSS-REFERENCES RIGHT.
ALL MACROS ARE INCLUDED.

\documentstyle[11pt]{article}
\catcode`@=11
\def\fo{\hbox{{1}\kern-.25em\hbox{l}}}

\def\ch{\@startsection{section}{1}{\z@}{-3ex plus-1ex minus-.2ex}%
        {2ex plus.2ex}{\large\sc}}

\def\ppm{pseudoparticle mechanical }
\def\ospn2{osp($N |$ 2)}

\def\ddt#1{\frac{d \;}{dt} (#1)}
\def\epsab{\epsilon_{\alpha \beta}}

\def\7#1#2{\mathop{\null#2}\limits^{#1}}        
\def\5#1#2{\mathop{\null#2}\limits_{#1}}        

\def\dddot#1{\hbox{$\mathop{#1}\limits^{\ldots}$}}

\newcommand{\NPB}[3]{{\sl Nucl. Phys.} {\bf B#1} (19#2)  {#3}}
\newcommand{\PRD}[3]{{\sl Phys. Rev.} {\bf D#1} (19#2)   {#3}}
\newcommand{\PLB}[3]{{\sl Phys. Lett.} {\bf #1B} (19#2)  {#3}}

\newcommand{\IJA}[3]{{\sl Int. Journ. of Mod. Phys.} {\bf A#1} (19#2) {#3}}
\newcommand{\CMP}[3]{{\sl Commun. Math. Phys.} {\bf #1} (19#2) {#3}}
\newcommand{\JMP}[3]{{\sl J. Math. Phys.} {\bf #1} (19#2) {#3}}

\newcommand{\PRep}[3]{{\sl Phys. Rep.} {\bf #1} (19#2)  {#3}}
\newcommand{\CQG}[3]{{\sl Class. Quant. Grav.} {\bf #1} (19#2)  {#3}}
\newcommand{\PTP}[3]{{\sl Prog. Theor. Phys.} {\bf #1} (19#2)  {#3}}
\newcommand{\TMP}[3]{{\sl Th. Math. Phys.} {\bf #1} (19#2)  {#3}}
\newcommand{\ZPC}[3]{{\sl Z. Phys.} {\bf C#1} (19#2)  {#3}}

\def\secteqno{\@addtoreset{equation}{section}%
\def\theequation{\thesection.\arabic{equation}}}
\def\endsecteqno{\def\theequation{\@ifundefined{chapter}%
{\arabic{equation}}{\thechapter.\arabic{equation}}}}
\newcounter{subequation}
\def\thesubequation{\alph{subequation}}
\def\sneqnarray{\stepcounter{equation}\let\@currentlabel=\theequation
\setcounter{subequation}{1}
\def\@eqnnum{{\rm (\theequation\thesubequation)}}
\global\@eqcnt\z@\tabskip\@centering\let\\=\@eqncr\let\@@eqncr=\@@sneqncr
$$\halign to \displaywidth\bgroup\@eqnsel\hskip\@centering
 $\displaystyle\tabskip\z@{##}$&\global\@eqcnt\@ne
 \hskip 2\arraycolsep \hfil${##}$\hfil
 &\global\@eqcnt\tw@ \hskip 2\arraycolsep $\displaystyle\tabskip\z@{##}$\hfil
  \tabskip\@centering&\llap{##}\tabskip\z@\cr}
\def\endsneqnarray{\@@sneqncr\egroup $$\global\@ignoretrue}
\def\@@sneqncr{\let\@tempa\relax
   \ifcase\@eqcnt \def\@tempa{& & &}\or \def\@tempa{& &}
   \else \def\@tempa{&}\fi
     \@tempa \if@eqnsw\@eqnnum\stepcounter{subequation}\fi
     \global\@eqnswtrue\global\@eqcnt\z@\cr}

\catcode`@=12

\def\beq{\begin{equation}}
\def\eeq{\end{equation}}
\def\beqar{\begin{eqnarray}}
\def\eeqar{\end{eqnarray}}
\def\beqarn{\begin{eqnarray*}}
\def\eeqarn{\end{eqnarray*}}
\def\half{\frac{1}{2}}
\def\vep{\varepsilon}

\secteqno

\title{\bf Superconformal theories from Pseudoparticle Mechanics}

\author{{\sc Karyn M. Apfeldorf},  {\sc Joaquim Gomis}%
\thanks{On leave of absence from Dept.\ d'Estructura i
Constituents de la Mat\`eria, U. Barcelona.} \\
\llap%
\small{\it Theory Group, Department of Physics} \\
\small{\it University of Texas at Austin } \\
\small{\it RLM\,5.208, Austin, TEXAS }  \\
{\it e-mail:} \small{apfel@utpapa.ph.utexas.edu},
              \small{gomis@utaphy.ph.utexas.edu}   }

\date{}

\begin{document}
\maketitle
\thispagestyle{empty}

\begin{abstract}
We consider a one-dimensional Osp($N|2M$) pseudoparticle mechanical
model which may be written as a phase space gauge theory.
We show how the pseudoparticle model naturally encodes and explains the
two-dimensional zero curvature approach to finding extended conformal
symmetries.   We describe a procedure of partial gauge fixing of these
theories which leads generally to theories with superconformally extended
${\cal W}$-algebras.
The pseudoparticle model allows one to derive the finite
transformations of the gauge and matter fields occurring in these
theories with extended conformal symmetries.
In particular, the partial gauge fixing of the Osp($N|2$) pseudoparticle
mechanical models results in theories with the
SO($N$) invariant $N$-extended superconformal symmetry algebra of
Bershadsky and Knizhnik.  These algebras are nonlinear for $N \geq 3.$
We discuss in detail the cases of $N=1$ and $N=2,$ giving
two new derivations of the superschwarzian derivatives.
Some comments are made in the
$N=2$ case on how twisted and topological theories represent a
significant deformation of the original particle model.
The particle model also allows one to interpret superconformal
transformations as deformations of flags in super jet bundles over
the associated super Riemann surface.
\end{abstract}

\vfill\hfill
\vbox{
\hfill March 1993 \null\par
\hfill UTTG-93-001}
\newpage

\section{Introduction}

Extended conformal symmetries play a central role in
many two-dimensional systems including string theories,
two-dimensional gravity theories, statistical mechanical
systems at phase transition points, as well as in integrable
hierarchies of nonlinear differential equations.

There has been a great deal of research on the
systematic construction and understanding of
bosonic and/or fermionic extensions of the two-dimensional
conformal symmetry algebra, or Virasoro algebra.
As far back as 1976, Ademollo, et.al.~\cite{ademollo}, generalized the
two-dimensional $N=1$ superconformal algebra to $N$-extended
O($N$) superconformal algebras, for any $N.$
These algebras are graded Lie algebras with a spin 2 stress
tensor, $N$ spin $\frac{3}{2}$ supersymmetry currents,
$N(N-1)/2$ spin 1 O($N$) Kac-Moody currents, and
$2^N -1 - N(N+1)/2$ additional generators.

An explosion of progress occurred after the paper of
Belavin, Polyakov and Zamolodchikov~\cite{BPZ} in which a method was
developed for studying  conformal symmetry through
operator product expansions (OPE's) and conformal Ward identities.
The semi-direct product of Virasoro with a Kac-Moody
algebra of spin 1 currents was investigated in~\cite{KZ}, and
more generally the systematic extension of Virasoro by bosonic
currents was initiated by Zamolodchikov and by Zamoldchikov
and Fateev~\cite{Zam}.  This program gives rise to the ${\cal W}$-algebras.
For recent reviews on ${\cal W}$-algebras, see~\cite{feher,bouwk}.

Supersymmetric extensions were considered by Knizhnik~\cite{kniznik} and
by Bershadsky~\cite{bersh} from the OPE point of view, and the resulting
algebras have important differences from those of Ademollo.
In particular, Bershadsky and Knizhnik
found SO($N$) and U($N$) invariant $N$-extended superconformal
algebras containing only spin $\frac{3}{2} $ and an SO($N$) or
U($N$) of spin 1 currents in addition to the spin 2 energy-momentum tensor.
These algebras are not graded Lie algebras for $N \geq 3,$ since
the OPE of two spin $\frac{3}{2}$ supercurrents yields
a term bilinear in the Kac Moody currents.  In the case of
the SO($N$) invariant algebras, one has a spin 2 energy-momentum
tensor, $N$ spin $\frac{3}{2}$ fermionic stress tensors and
$N(N-1)/2$ spin 1 currents forming a SO($N$) current algebra.

A common feature of the ${\cal W}$-algebras and the
$N \geq 3$ Bershadsky-Knizhnik SO($N$) and U($N$) superconformal
algebras is that their OPE's are nonlinear, and
therefore Lie algebraic techniques are not directly applicable.
Two methods have enjoyed much success.
Quantum hamiltonian reduction,
the quantum version of the Drinfel'd-Sokolov reduction
for coadjoint orbits,  furthered understanding of
extended conformal algebras by showing that they
may be obtained from constrained Kac Moody current algebras.
The Bershadsky-Knizhnik algebras may be obtained from reduction
of Osp($N|2$) current algebras~\footnote{For a review and
list of references, see~\cite{imp}.}.
Another method is the zero curvature approach~\cite{polyakov,bilal,das}
which gives a prescription for
determining the infinitesimal transformations of
gauge fields of extended conformal algebras from
two-dimensional gauge theories, provided one identifies
a spacetime derivative as the gauge variation operation.
While these approaches have provided insight into extended
conformal algebras, a complete understanding of the geometry
associated with these symmetries is still lacking.
A major shortcoming of quantum hamiltonian reduction and the zero
curvature method is that they give only OPE's of the
reduced algebras, or equivalently determine only the
infinitesimal transformations of the symmetries.
On the other hand, progress has been made in understanding
${\cal W}$-geometry in the work of Gerasimov, Levin, Marshakov~\cite{geralm},
and of Bilal, et. al.~\cite{bilal}, where it is shown that ${\cal W}$
transformations may be regarded as deformations
of flags in jet bundles over the two-dimensional Riemann surfaces.
Other approaches to ${\cal W}$-geometry have been considered
by~\cite{deBoer,schout,sotkov,hull,diFran,gervais,fof}.

In this paper, we present one-dimensional
pseudoparticle mechanical models which may be written as
Osp($N|2M$) phase space gauge theories.  The partial gauge fixing
of these theories results in theories with superconformal
${\cal W}$-algebras, and the resulting theories may be considered
as chiral sectors of two-dimensional super ${\cal W}$-gravity theories.
The particle model formulated as a gauge theory in phase space
sheds light on the two-dimensional zero curvature prescription.
Of particular interest are the Osp($N|2$) pseudoparticle mechanical models
whose partial gauge fixing
results in the SO($N$)-invariant $N$-extended superconformal
theories of Bershadsky and Knizhnik.   Upon partial gauge fixing, one
obtains chiral sectors of two-dimensional theories of conformal matter
coupled to supergravity.
This reduction is fundamentally different than the
hamiltonian reduction of Wess-Zumino-Novikov-Witten (WZNW) models,
as the particle model contains both gauge and matter fields which possess only
canonical Dirac brackets, as opposed to the WZNW model which
possesses only gauge currents with the Lie-Poisson Kac-Moody brackets.
A method for obtaining the finite transformations for
SO($N$) extended superconformal algebras, valid also for
${\cal W}$-extensions of the these algebras, is presented.
The nonsupersymmetric case is studied in~\cite{gomishkr}.
In brief, the method is as follows.
First one makes gauge field dependent redefinitions of the gauge parameters
to put the infinitesimal Osp($N|2$) transformations of the matter and
gauge fields into ``standard'' form, i.e. the form in which one may
immediately recognize diffeomorphisms and supersymmetries.
This step is of practical necessity since otherwise it would be difficult
to recognize, for example, ordinary diffeomorphisms
in the algebra after partial gauge fixing.   At this stage the choice
of partial gauge fixing condition becomes clear.
Next one integrates the infinitesimal Osp($N|2$) transformations and
transforms the matter and gauge fields by successive finite transformations.
Finally, one imposes the partial gauge fixing at the level of the finite
transformations.  This yields the finite transformations for the
matter fields and non-gauge-fixed gauge fields.
This method is illustrated by explicit calculations for the
(albeit linear) cases of $N=1$ and $N=2.$
The pseudoparticle model also facilitates the interpretation of superconformal
transformations as deformations of flags in the super jet
bundles over the associated super Riemann surfaces.

This paper is organized as follows.
In section 2, we illustrate how an Osp($N|2M$) pseudoparticle
mechanical model can be formulated as a phase space gauge theory,
and explain the connection with the zero curvature prescription.
We comment on the partial gauge fixing procedure from the
point of view of orbits of the group, and contrast the reduction
of the pseudoparticle model with the hamiltonian reduction of a WZNW model.
In section 3, we restrict our considerations to the Osp($N|2$)
model.  We discuss the general method for obtaining the finite
transformations for SO($N$)-invariant  $N$-extended superconformal
algebras.  We discuss the closure of the infinitesimal gauge algebra
and illustrate the partial gauge fixing of the model at the
infinitesimal level explicitly for $N=1$ and $N=2.$
In section 4, we carry out the gauge fixing at the level of
finite transformations for $N=1.$  Finite transformations
are given for matter and gauge fields, thus providing a derivation
of the $N=1$ superschwarzian.
In section 5, we present the cases of $N=1$ and $N=2$ in
superfield form.   Here we give an alternate derivation of the
superschwarzians by writing the matter equation of motions in
superfield form and then demanding the covariance of the
equation of motion under superconformal transformations.
In section 6, we discuss the completely gauge fixed  pseudoparticle
model, which is invariant under super M\"{o}bius transformations.
In section 7, we return to the infinitesimal $N=2$ and discuss
how twisted and topological theories represent a significant
deformation of the original particle model.
In section 8, we show how superconformal
transformations may be understood as deformations of flags in
the $N$-supersymmetrized 1-jet bundles over the super Riemann surfaces.
We construct the  flags in the super 1-jet bundle explicitly
for $N=1$ and for $N=2,$ and conjecture the result for general $N.$
Section 9 contains the conclusions and some directions
for further investigation.

\section{Osp($N | 2M$) Pseudoparticle Mechanical
Models as Phase Space Gauge Theories}

Consider a one-dimensional hamiltonian system with
$M$ bosonic and $N$ fermionic dynamical coordinates
\beqar
( x_i(t)^\mu, p_i(t))_\mu) & & \quad \quad i=1,\ldots M \\
( \psi_\alpha(t)^\mu, \pi_{\psi_\alpha}(t)_\mu)
& & \quad \quad \alpha =1,\ldots N  \nonumber
\eeqar
and einbein-like coordinates
\beq
(\lambda_A(t),\pi_{\lambda_A}(t))
\eeq
which will act as Lagrange multipliers to implement a set of
constraints $T_A(x,p,\psi)$ on the dynamical phase
space coordinates.
The index $\mu$ runs over the $d$ spacetime dimensions, where
$d$ is sufficiently large so the constraints do not collapse the theory.
The metric $g_{\mu \nu}$ should be taken not as euclidean, but
as diagonal 1's and -1's.

The canonical action is
(summation convention assumed and spacetime indices suppressed)
\beq
S = \int \: dt \left[ \dot{x}_i \cdot p_i + \half \dot{\psi}_\alpha
\cdot \psi_\alpha - \lambda_A \; T_A(x,p,\psi)   \right] .
\eeq
where the fermion momenta $\pi_{\psi_\alpha}$ have been eliminated by the
second class constraints
$\pi_{\psi_\alpha} - \frac{1}{2} \psi_\alpha = 0$.
The fundamental Dirac-Poisson brackets of this model are
\beq
\{x^\mu_i ,p_{\nu \:j} \}^*  = g^\mu_{\; \nu} \delta_{ij} \quad \quad
\{ \psi_\alpha^\mu, \psi_\beta^\nu \}^*
= - \delta_{\alpha \beta} \: g^{\mu \nu}.
\eeq
This system has the primary constraints
$
\pi_{\lambda_A} \approx 0
$
whose stability implies the secondary constraints
$
T^A(x,p,\psi)   \approx  0 .
$
In turn, the stability of the constraints $T_A \approx 0$
implies that (in suggestive notation)
$$
\{ T_i, T_j \} = f_{ij}^{\:\:\:k} T_k \approx 0
$$
where the $f_{ij}^{\:\:\:k}$ could in general depend on the coordinates
$(x,p,\psi).$

Now let us examine a system where the constraints are
all quadratic combinations of the coordinates $x_i, p_i,$ and
$\psi_\alpha$
\beq
\begin{array}{llll}
T_{1ij} = \half p_i p_j \quad & T_{2ij} = - \half x_i x_j  \quad
& T_{3ij} = \half p_i x_j \quad & (i \leq j) \\
 & & \\
T_{4 i \alpha} = \half p_i \psi_\alpha  \quad & T_{5 i \alpha} =
\half x_i \psi_\alpha  \quad & T_{6 \alpha \beta}
= \frac{1}{4} \psi_\alpha \psi_\beta \quad & (\alpha < \beta) .
\end{array}
\eeq

\noindent
Using the fundamental Dirac-Poisson brackets of this theory
one can check that the Poisson algebra formed
by the constraints $T_A$ is isomorphic to the Lie algebra
osp($N | 2 M$).

The action of this theory is
\beqar
S &=& \int \: dt \left[ \dot{x_i} \cdot p_i
+ \half \dot{\psi}_\alpha \cdot \psi_\alpha
- (\lambda_1)_{ij} \frac{p_i p_j}{2} + (\lambda_2)_{ij} \frac{x_ix_j}{2}
\right.  \\ \nonumber & &  \left.
- (\lambda_3)_{ij} \frac{x_i p_j}{2}
- (\lambda_4)_{i \alpha} \frac{p_i \psi_\alpha}{2}
- (\lambda_5)_{i \alpha} \frac{x_i \psi_\alpha}{2}
- (\lambda_6)_{\alpha \beta} \frac{\psi_\alpha \psi_\beta}{4} \right]  .
\eeqar

\noindent
Related particle models have been investigated
by~\cite{marnelius,howePPT,siegel,martensson,ikemori}.

Following the insights of Kamimura~\cite{kamimura} on
reparametrization invariant theories,
this model may be understood better by using matrix notation.
In particular, we can put the action into standard Yang Mills form,
with gauge group Osp($N | 2 M$).
Let us write the first order formalism coordinates  in an
$N+2M$ vector $\Phi.$
It will be convenient to write also the bosonic and fermionic
coordinates separately in $2M$ and $N$-dimensional vectors
\beq
\Phi =\left(\begin{array}{c}
        \phi \\
        \psi
       \end{array}\right),\quad\quad
\phi =\left(\begin{array}{c}
        x_1\\
        \vdots \\
  	x_M \\
	p_1 \\
	\vdots_M \\
	p_M
       \end{array}\right),\quad\quad
\psi =\left(\begin{array}{c}
        \psi_1 \\
	\vdots\\
	\psi_N
        \end{array}\right).
\eeq
The orthosymplectic group Osp($N | 2 M $) consists of those elements
leaving fixed the quadratic form
$$ \xi^\top \eta \xi  \quad \quad
\hbox{where      }
\eta = \left(\begin{array}{cc}
        J_{2M} & 0 \\
	0 & 1_N
	\end{array} \right)
$$
where $J_{2M}$ is the $2M \times 2M$ symplectic matrix
$
J_{2M} = \left(\begin{array}{cc}
        0 & 1_M \\
	-1_M & 0
	\end{array} \right)
$
and $1_N$ is the $ N \times N $ unit matrix.
The conjugate to $\Phi$ is therefore given by
$$
\bar \Phi =  \Phi^\top \eta = \left( \phi^\top , \psi^\top \right) \eta
= \left(- p_1,\ldots,-p_M,x_1,\dots,x_M,\psi_1,\ldots,\psi_N \right).
$$
The Lagrange multiplier gauge fields may be assembled into the
form of a most general osp($N|2 M$) matrix~\footnote{Recall the
transpose of a supermatrix with bosonic blocks $b_1$ and $b_2$ and
fermionic blocks $f_1$ and $f_2$ is given by
$$ \left(\begin{array}{cc}
b_1 & f_1 \\ f_2 & b_2 	\end{array} \right)^\top  =
\left(\begin{array}{cc}
b_1^\top & f_2^\top \\ - f_1^\top & b_2^\top \end{array} \right).
$$}
\beq
\Lambda = \left(\begin{array}{cc}
        A & \Omega \\
	\Omega^\top J_{2M} & B
	\end{array} \right)
\eeq
where $A$ is a $2M \times 2M$ symplectic matrix,
$\Omega$ is a $N \times 2M$ matrix of fermionic entries, and
$B$ is an $N \times N$ antisymmetric matrix.
The matrices $A,$    $\Omega$ and $B$  may be written explicitly in terms
of the gauge fields.   The matrix $A$ is an sp(2$M$) matrix
\beq
A  = \left(\begin{array}{cc}
       \half \lambda_3  & \lambda_1 \\
        \lambda_2 & - \half \lambda_3^\top
        \end{array}\right)
\eeq
where the components of the $M \times M$ matrices
$(\lambda_1)_{ij},(\lambda_2)_{ij}, (\lambda_3)_{ij}$
are the Lagrange multiplier fields appearing in the action above.
The fermionic gauge fields are arranged as
\beq
\Omega = \half \left(\begin{array}{c}
        \lambda_4  \\
      - \lambda_5
        \end{array}\right),
\eeq
where the $M \times N$ matrices
$(\lambda_4)_{i \alpha}$ and $(\lambda_5)_{i \alpha}$
appear in the action.  Finally, $B$ is the $N \times N$
antisymmetric matrix of Lagrange multipliers
\beq
B_{\alpha \beta} = - \half (\lambda_6)_{\alpha \beta},
\eeq
which implement the O($N$) rotations among the Grassmann variables.

Using this matrix notation, the canonical action of the constrained
pseudoparticle model may be written simply as a phase space gauge
theory
\beq
S= - \half \int \; dt \: \bar \Phi {\cal D} \Phi  ,
\eeq
where ${\cal D}$ is the covariant derivative
$$
{\cal D} = \frac{d \:}{dt} - \Lambda .
$$
The constrained pseudoparticle action is just an Osp($N | 2 M$)
gauge theory with no kinetic term for the Yang Mills gauge fields.

We are now in the position to make the connection with the
zero curvature method.
In this prescription, one considers a two dimensional theory
with gauge fields $A_z = A_z(z,\bar z),$ $A_{\bar{z}} =
A_{\bar{z}}(z,\bar z).$  Making a gauge fixing
ans\"{a}tz for $A_z$ and keeping the form of $A_{\bar{z}}$ general,
one solves the zero curvature equation, i.e.
$$
F_{z \bar{z}} = \left[ \partial-A_z, \bar{\partial}-A_{\bar{z}} \right]
\equiv 0
$$
to eliminate some elements of $A_{\bar{z}}$ and to determine
the $\bar \partial$ derivatives of the non-gauge-fixed components of
$A_z.$   If one makes the identification
$$ \bar{\partial} \leftrightarrow \delta $$
then, given a suitable ans\"{a}tz for $A_z,$ the resulting equations
for the nonconstrained elements give their infinitesimal transformations
under the residual extended conformal symmetry.

Let us now return to the pseudoparticle phase space gauge theory.
Under infinitesimal osp($N|2 M$) gauge transformations, the matter
and gauge fields transform in the usual way
\beqar
\delta_\epsilon \Phi  &=&  \epsilon  \Phi \\  \nonumber
\delta_\epsilon \Lambda &=& \dot \epsilon - \left[ \Lambda, \epsilon
\right],
\eeqar
where $\epsilon$ is an osp($N| 2M$) matrix of gauge parameters.
The equation of motion for the matter fields is
\beq
{\cal D} \Phi = \dot \Phi - \Lambda \Phi =0.
\eeq
The compatibility condition of the two linear equations
for the matter fields is nothing but the gauge transformation
equation for the gauge fields.  Explicitly, the compatibility
equation is
\beq
0 = \left[ \delta_\epsilon - \epsilon ,
\frac{d \:}{dt} - \Lambda \right] \Phi
 =  \underbrace{\left( -\dot{\epsilon} + \delta_\epsilon \Lambda +
\left[ \Lambda,\epsilon \right] \right) }_{\hbox{gauge field transformation}}
\Phi .
\eeq
This relation continues to hold when a partial
gauge fixing condition is imposed.
Thus, the one dimensional particle model allows one to make sense of the
the two-dimensional zero curvature approach.  The infinitesimal gauge
transformation equation and equation of motion for the matter in the model
provide the two linear operators $\partial - A_z$ and $\bar \partial -
A_{\bar{z}}$ appearing in the zero curvature approach, while the
gauge field transformation equation is equivalent to $F_{z \bar{z}}=0.$
{}From the particle model point of view there is no need to make the
curious identification $ \bar{\partial} \leftrightarrow \delta $.
{}From the point of view of the particle model, the zero curvature
condition translates into a statement of the gauge symmetry and the
identification $ \bar{\partial} \leftrightarrow \delta$
is not required.

Before closing this section, we comment on the partial gauge fixing
of this Yang Mills type action.  The finite gauge transformations are
\beqar
\Phi^\prime &=& g \Phi   \\  \nonumber
\Lambda^\prime &=& g^{-1} \Lambda g - g^{-1} \dot{g} = Ad^*_g \Lambda
\label{eq:orbit}
\eeqar
where $g \in$ Osp($N| 2 M$).   The gauge transformed fields
$\Lambda^\prime$ sweep out the coadjoint orbit of the point $\Lambda.$
Let $\Lambda_c$ be the gauge field matrix with some entries
constrained to be ones or zeros.
If we demand that the gauge transformed matrices
$\Lambda_c^\prime$ leave these constraints intact,  then the
reduced orbit will describe finite transformations of the gauge
fields under the residual gauge symmetry.
In general the relations among the parameters will be gauge
field dependent and the symmetry will be a quasigroup~\cite{batalin}.
With a suitable choice of partial gauge fixing, the resulting
equations for the unconstrained elements of $\Lambda_c$ give finite
transformations for the gauge fields under extended conformal algebras.

Before discussing a technical obstruction to obtaining the finite
transformations directly from the group orbit, we contrast the
reduction of the pseudoparticle model with the corresponding
hamiltonian reduction of a WZNW model.
In the latter case, the original phase space that one reduces
is that of a chiral sector of WZNW currents.
A point in the phase space is specified by $(J(z),k)$
where $J(z) = J^a(z) T_a $ is a mapping from the circle to the
algebra (which is \ospn2 for the Bershadsky-Knizhnik superconformal
algebras) and $k$ is a number.   The $J^a(z)$ have Poisson brackets
which are isomorphic to the affine Lie algebra.
The coadjoint action on a phase space point is
$ Ad^*_g \left( J(z),k \right) = \left( g^{-1} J(z) g + k g^{-1}
\frac{d \;}{dz} , k \right) $.
In hamiltonian reduction, or the Drinfel'd-Sokolov reduction for
coadjoint orbits, one constrains some of the $J^a(z)$'s
(generally fewer entries are constrained than in the relevant
pseudoparticle model reduction),
and then mods out by the coadjoint action of the subgroup which
preserves these constraints.
The reduced orbits are then parametrized by new phase space
coordinates which are polynomials of the original phase space
coordinates and their derivatives.  In contrast to the pseudoparticle
model, the orbit equation is used to determine the new coordinates
instead of the transformations.  The original Poisson brackets are then used
to obtain the Poisson brackets of the extended conformal symmetry.
Thus, the reduction of the pseudoparticle model is markedly different
from the hamiltonian reduction of WZNW in two major ways.
The former includes both matter and Lagrange multiplier gauge fields
while the latter has gauge fields only with affine Lie-Poisson brackets.
The former employs the one-dimensional gauge theory coadjoint orbit
to determine the finite transformations, while the latter uses the
affine coadjoint action to determine the new phase space coordinates.

Returning to the pseudoparticle model, consider the residual
gauge transformations arising from the gauge orbit equation~\ref{eq:orbit}
by directly plugging in constraints on $\Lambda.$
Unfortunately, the transformations obtained in this manner will not
be directly recognizable as  corresponding to extended conformal algebras
because the gauge parameters would not correspond in a simple way
to ``standard'' transformations such as diffeomorphisms or
supersymmetries for example.  To put expressions in a recognizable
form would require gauge field dependent parameter redefinitions
which from a practical point of view is unfeasible.
Instead we will take a different approach, whose first step involves making
field redefinitions first at the infinitesimal level to get
``standard'' transformations.   For arbitrary $N$ and
$M=1$ there is no appreciable
difficulty in finding the standard  form of the transformations,
whereas for $M \geq 2$ where the ${\cal W}$-algebras appear, the
standard transformations are not so well established.
The nonsupersymmetric case has been investigated in~\cite{gomishkr}.
In the remainder of this paper, we will restrict to $M=1$
to focus on the $N$-extended superconformal algebras, which are
nonlinear for $N \geq 3$.

\section{Superconformal theories from osp($N|$2) pseudoparticle model}
\noindent

We now restrict to \ppm models which have as dynamical coordinates
a single bosonic coordinate $x^\mu(t)$ and its momentum $p_\mu(t)$
and $N$ fermionic coordinates $ \psi_\alpha^\mu(t) $
with $\alpha = 1, \ldots N.$  The constraints
\beq
\begin{array}{llll}
T_1 = \half p^2 \quad & T_2 = - \half x^2  \quad & T_3 = \half p x &  \\
 & & \\
T_{4 \alpha} = \half p \psi_\alpha  \quad & T_{5 \alpha} =
\half x \psi_\alpha  \quad & T_{6 \alpha \beta} = \frac{1}{4}
\psi_\alpha \psi_\beta  \quad & (\alpha < \beta)
\end{array}
\eeq
form an \ospn2 algebra, with $T_1,$ $T_2$ and $T_3$ forming
an sl(2,R) subalgebra.  We pass to the lagrangian form, integrating
out the momentum
\beqar
S &=& \int \: dt \left[ \frac{1}{2 \lambda_1}
(\dot{x} - \half \lambda_3 x - \half \lambda_{4 \alpha} \psi_\alpha )^2
+  \frac{1}{2} \dot{\psi}_\alpha \psi_\alpha  + \lambda_2 \frac{x^2}{2}
\right.   \nonumber \\
& & \left.  - \lambda_{5 \alpha} \frac{x \psi_\alpha}{2}
- \lambda_{6 \alpha \beta}  \frac{\psi_\alpha \psi_\beta}{4} \right] .
\eeqar
This action describes a relativistic spinning particle
moving in a $d$-dimensional spacetime and interacting with
gauge fields $\lambda_i$ which subject the particle to an
\ospn2 algebra of constraints.  This action has been
considered by M{\aa}rtensson~\cite{martensson}.
Denoting the pullback of the momentum by
\beq
\kappa \equiv FL^* p = \frac{1}{\lambda_1} ( \dot{x}
- \half \lambda_3 x - \half \lambda_{4 \alpha} \psi_\alpha )
\label{eq:pullback}
\eeq
the equations of motion are
\beqar
S_x &=& \lambda_2 x - \half \lambda_3 \kappa - \half \lambda_{5
\alpha} \psi_\alpha - \dot{\kappa} \nonumber \\
S_{\psi_\alpha} &=& - \dot{\psi}_\alpha + \half \kappa \lambda_{4 \alpha}
+ \half x \lambda_{5 \alpha}
- \half \lambda_{6 \alpha \beta} \psi_\beta.
\eeqar

Our method for obtaining $N$-extended superconformal theories
from the pseudoparticle models is composed of the following steps:
\begin{enumerate}
\item Put Osp($N|2$) infinitesimal transformations in ``standard''
form by gauge field dependent redefinitions of gauge parameters.
Determine the partial gauge fixing condition.
\item Integrate the linear \ospn2 algebra to get the finite
transformations, and transform the matter and gauge fields by
successive finite transformations.
\item Perform partial gauge fixing at the finite level, thus
obtaining finite transformations for $N$-extended
superconformal transformations.
\end{enumerate}

This is a very general prescription which will work for particle
models with linear gauge symmetries. The partial gauge fixing of
Osp($N|2M$) will in general result in superconformal ${\cal W}$-algebras.
It is unwise to attempt to reverse the order of integration
and partial gauge fixing since in general the residual infinitesimal
gauge symmetries will be nonlinear and therefore difficult or
impossible to integrate.
Partial gauge fixing and integration do \underline{not} in general commute,
and therefore will not in general produce the same finite
transformations for the residual symmetries of the gauge
fixed model (the transformations will be related by gauge field
dependent redefinitions of parameters).
This is clear is we consider the orbit equation that we are solving.
If we do a partial gauge fixing $\Lambda_c$ at the infinitesimal level,
then solving $0=\delta \Lambda = ad_\epsilon^* \Lambda $
implies some gauge field dependent relations along the
gauge parameters, giving a matrix $\epsilon_c$.  Integrating this
gives $\Lambda_c^\prime = Ad_{e^{\epsilon_c}}^* \Lambda_c$ for the
finite transformations.  On the other hand, partial gauge fixing at the
finite level means that we solve
$\Lambda_c^\prime = Ad_g^* \Lambda_c,$ which will give in general
different gauge field dependent relations among the gauge parameters.

If we work in first order formalism, the gauge algebra on the
phase space coordinates $x,$ $p,$ $\psi_\alpha$ and on the
gauge fields $\lambda_A$ closes, i.e.
$ \left[ \delta_{\varepsilon_1},\delta_{\varepsilon_2} \right]
= \delta_{\varepsilon_*}  .$
If we instead work in the Lagrangian formalism, then the
transformation is the pullback of the former transformation,
and generally one can expect the equations of motions of the
coordinates to appear in the commutator of the gauge transformations.
The transformations of the gauge fields are unaffected since $p$ does not
enter into the expressions.
Explicitly, the gauge variations of the coordinates are
\beqar
\delta_\vep x  &=& \vep_1 \kappa + \half \vep_3 x
+ \half \vep_{4 \alpha} \psi_\alpha  \label{eq:xtrans} \\
\delta_{\vep} \psi_\alpha &=& \half \vep_{4 \alpha} \kappa +
\half \vep_{5\alpha} x - \half \vep_{6 \alpha \beta} \psi_\beta .
\eeqar
The pullback of the gauge variation of the momentum and
the gauge variation of the pullback of the momentum do not coincide
\beqar
FL^* \delta_\vep p &=& \vep_2 x  - \half \vep_3 \kappa - \half \vep_{5 \alpha}
\psi_\alpha  \nonumber \\
\delta_\vep FL^* p &=& \vep_2 x  - \half \vep_3 \kappa - \half \vep_{5 \alpha}
\psi_\alpha  - \vep_1 \frac{S_x}{\lambda_1}
- \vep_{4 \alpha} \frac{S_{\psi_\alpha}}{2 \lambda_1}.
\eeqar
The noncoincidence of the momentum and its pullback is
an indication that the gauge algebra in the Lagrangian
formulation may not be closed.  A calculation reveals
\beqar
\left[ \delta_\eta,\delta_\vep \right] x &=& \delta_{\left[ \vep,\eta \right]
} x
+ (\vep_1 \eta_{4 \alpha} - \eta_1 \vep_{4 \alpha})
\frac{- S_{\psi_\alpha}}{2 \lambda_1} \\
\left[ \delta_\eta,\delta_\vep \right] \psi_\alpha &=&
\delta_{ \left[ \vep,\eta \right] }
\psi_\alpha + (\vep_1 \eta_{4 \alpha} - \eta_1 \vep_{4 \alpha})
\frac{ S_x}{2 \lambda_1}  + (\eta_{4 \beta} \vep_{4 \alpha}
+ \eta_{4 \alpha}  \vep_{4 \beta}) \frac{S_{\psi_\beta}}{4
\lambda_1} \nonumber .
\eeqar

At this point, it is useful to change from the parameters occurring
naturally in the gauge theory to parameters which represent
``standard'' transformations.   In the following formulas, we
now restrict further to the case of $N \leq 2$. For $N=2,$ we write the
single $\lambda_{6 \alpha \beta}$ field as  $\lambda_6$ and its
associated parameter as $\epsilon_6.$  These are otherwise ($N=0,1$) absent.
The procedure for arbitrary $N$ is carried out in exactly the same manner,
with the inclusion of a total of $\frac{N(N-1)}{2}$  $\lambda_{6 \alpha
\beta}$  gauge fields and their associated gauge parameters.
Let us make the following change of variables
\beqar
\vep_1 &=& \lambda_1 \xi  \nonumber \\
\vep_2 &=& \lambda_2 \xi + h  \nonumber \\
\vep_3 &=& \lambda_3 \xi + \sigma \nonumber \\
\vep_{4 \alpha} &=& \lambda_{4 \alpha} \xi + 2 \omega_\alpha
\nonumber \\
\vep_{5 \alpha} &=& \lambda_{5 \alpha} \xi + \chi_\alpha  \nonumber \\
\vep_6 &=& \lambda_6 \xi + R,
\eeqar
where the only real modification is in the reparametrization.
By inspection it is clear that the transformations are as follows:
reparametrization $\xi,$ scale $\sigma,$ shift transformation $h,$
supersymmetries $\omega_\alpha,$ further fermionic symmetries $\chi_\alpha,$
and in the case of $N=2,$ we have O(2) rotations parametrized by $R$.

Upon examination of $\delta \psi_\alpha,$ we see that to
get standard reparametrization, we must add a term to the
variation of $\psi_\alpha$
$$
\delta \psi_\alpha \rightarrow \delta \psi_\alpha - \xi S_{\psi_\alpha}.
$$
\noindent
The infinitesimal forms of the gauge transformations are
\beqar
\delta x &=& \xi \dot{x} + \half \sigma x + \omega_\alpha \psi_\alpha
\nonumber \\
\delta \psi_\alpha &=& \xi \dot{\psi}_\alpha + \omega_\alpha \kappa
+ \half \chi_\alpha x - \half R \epsab \psi_\beta \nonumber \\
\delta \lambda_1 &=& \ddt{\xi \lambda_1}
+ \sigma \lambda_1 + \omega_\alpha \lambda_{4 \alpha} \nonumber \\
\delta \lambda_2 &=& \ddt{\xi \lambda_2} +
\dot{h} + h \lambda_3 - \sigma \lambda_2  - \half \chi_\alpha
\lambda_{5 \alpha} \nonumber \\
\delta \lambda_3 &=& \ddt{\xi \lambda_3}
+ \dot{\sigma} - 2 h \lambda_1 + \omega_\alpha \lambda_{5 \alpha}
+ \half \chi_\alpha \lambda_{4 \alpha} \nonumber \\
\delta \lambda_{4 \alpha} &=& \ddt{\xi \lambda_{4 \alpha} }
+ 2 \dot{\omega}_\alpha
+ \chi_\alpha \lambda_1 + \half \sigma \lambda_{4 \alpha}
- \omega_\alpha \lambda_3 + \epsab \omega \lambda_6 \nonumber  \\
& & - \half R \epsab \lambda_{4 \beta} \nonumber \\
\delta \lambda_{5 \alpha} &=& \ddt{\xi \lambda_{5 \alpha}}
- h \lambda_{4 \alpha}
+ 2 \omega_\alpha \lambda_2 - \half \sigma \lambda_{5 \alpha}
+ \half \chi_\alpha \lambda_3 + \half \epsab
\chi_\beta \lambda_6  \nonumber \\
&  & - \half R \epsab \lambda_{5 \beta} + \dot{\chi}_\alpha  \nonumber \\
\delta \lambda_6 &=& \xi \ddt{\xi \lambda_6 }
+ \dot{R} + \epsab \omega_\alpha \lambda_{5 \beta}
+ \half \epsab \chi_\beta \lambda_{4 \alpha}
\eeqar

\noindent
The gauge algebra for $x$ and for all the $\lambda_i$ closes
as $ \left[ \delta_A, \delta_B \right] = \delta_* $
with the starred parameters given below
\beqar
\xi^* &=& \xi^B \dot{\xi}^A - \xi^A \dot{\xi}^B - \frac{2}{\lambda_1}
\omega_\alpha^A \omega_\alpha^B  \nonumber \\
\sigma^* &=& \xi^B \dot{\sigma}^A - \xi^A \dot{\sigma}^B + \frac{2}{\lambda_1}
\omega_\alpha^A \omega_\alpha^B \lambda_3 + \chi_\alpha^B
\omega_\alpha^A  - \chi_\alpha^A \omega_\alpha^B \nonumber  \\
\omega_\alpha^* &=& \xi^B \dot{\omega}_\alpha^A - \xi^A \dot{\omega}_\alpha^B
+ \frac{1}{\lambda_1} \omega_\beta^A \omega_\beta^B \lambda_{4 \alpha}
+ \half (\sigma^B \omega_\alpha^A  - \sigma^A \omega_\alpha^B )
\nonumber  \\
& & + \half \epsab (R^A \omega_\beta^B - R^B
\omega_\beta^A)  \nonumber \\
h^* &=& \xi^B \dot{h}^A - \xi^A \dot{h}^B + \frac{2}{\lambda_1}
\omega_\alpha^A \omega_\alpha^B \lambda_2
+ h^B \sigma^A - h^A \sigma^A  + \half \chi_\alpha^A \chi_\alpha^B
\nonumber \\
\chi_\alpha^* &=& \xi^B \dot{\chi}_\alpha^A - \xi^A \dot{\chi}_\alpha^B
+ \frac{2}{\lambda_1} \omega_\beta^A \omega_\beta^B \lambda_{5 \alpha}
-2 (h^B \omega_\alpha^A  - h^A \omega_\alpha^B)   \nonumber \\
& & -\half ( \sigma^B \chi_\alpha^A - \sigma^A \chi_\alpha^B)
   + \half \epsab (R^A \chi_\beta^B - R^B \chi_\beta^A) \nonumber \\
R^* &=& \xi^B \dot{R}^A - \xi^A \dot{R}^B + \frac{2}{\lambda_1}
\omega_\alpha^A \omega_\alpha^B \lambda_6 +
\epsab (\omega_\alpha^B \chi_\beta^A - \omega_\alpha^A \chi_\beta^B) .
\eeqar

\noindent
The supersymmetry part of the gauge algebra on $\psi_\alpha$ is
still not closed for $N=2$ (and will not be for $N \geq 2$)
\beqar
[\delta_A, \delta_B ] \psi_\alpha &=&
\xi^* \dot{\psi}_\alpha + \omega_\alpha^* \kappa + \half \chi_\alpha^*
x + \half R^* \epsab \psi_\beta
- \frac{2}{\lambda_1} \omega_\beta^A \omega_\beta^B S_{\psi_\alpha}
\nonumber  \\
& & + \frac{1}{\lambda_1} (\omega_\beta^A \omega_\alpha^B
 - \omega_\beta^B \omega_\alpha^A ) S_{\psi_\beta}
\eeqar
In order to close the algebra for $N=2$ and still maintain the
the standard form of reparametrization for $\psi_\alpha,$
one must introduce an auxiliary field $F$ in the following way.
The transformation for $\psi_\alpha$  is further modified as
$$
\delta \psi_\alpha \rightarrow \delta \psi_\alpha + \epsilon_{\alpha
\beta } \omega_\beta F
$$
and $F$ must transform as
\beq
\delta F = \xi \dot{F} - \half \sigma F + \frac{1}{\lambda_1}
\epsab \omega_\alpha S_{\psi_\beta}
-\frac{1}{2 \lambda_1} \omega_\beta \lambda_{4 \beta} F.
\eeq
The action will be modified by a term
\beq
\delta S =  \half \int \; dt \; \lambda_1 \; F^2 .
\eeq
With this auxiliary field, the gauge algebra closes on $\psi_\alpha$
with the starred parameters listed above.
More generally, auxiliary fields should be added to exactly compensate
for the lack of noncommutativity of the pullback and the gauge
variation.  For example, by inspection
the case of $N=3$ requires three bosonic auxiliary fields to close the algebra.

With the infinitesimal transformations in standard form,
we are a position to determine the partial gauge fixing choice.
Reparametrizations should be generated by the $p^2/2$ term
in the lagrangian, and therefore the $\vep_1 \kappa $
term in $\delta_\vep x$ should give rise to reparametrizations of $x.$
{}From the equations~\ref{eq:xtrans} and~\ref{eq:pullback} (valid for
all $N$)
$$
\delta_\vep x  \supset \frac{\vep_1}{\lambda_1}
( \dot{x} - \half \lambda_3 x - \half \lambda_{4 \alpha} \psi_\alpha ),
$$
it is clear that for any $N$ the partial gauge fixing choice is
$\lambda_1 =1 ,$ $\lambda_3=0$ and $\lambda_{4 \alpha} =0 .$
Note that the number of gauge field degrees of freedom
is originally $3 + 2 N + N(N-1)/2.$  After imposing the
$2 + N$ gauge fixing conditions, we have $1 + N + N(N-1)/2$
gauge fields remaining, which is precisely the number we expect
for the SO($N$)-invariant $N$-extended superconformal algebra.
The gauge fixed lagrangian is
\beq
S = \int \: dt \left[ \half \dot{x}^2 +  \frac{1}{2}
\dot{\psi}_\alpha \psi_\alpha  + \lambda_2 \frac{x^2}{2}
- \lambda_{5 \alpha} \frac{x \psi_\alpha}{2}
- \lambda_6 \epsab  \frac{\psi_\alpha \psi_\beta}{4}
+ \half F^2 \right] .
\eeq
This manifestly conformally invariant lagrangian without the
auxiliary field has been studied by Siegel~\cite{siegel}.

While we will eventually do the gauge fixing at the finite level,
it is enlightening to consider the partial gauge fixing at the
infinitesimal level as well.
Imposing the constraints gives
\beq
\left\{  \begin{array}{rl}
\delta(\lambda_1 -1) & = 0 = \dot{\xi} + \sigma  \\
\delta \lambda_3     & = 0 = \dot{\sigma} - 2 h + \omega_\alpha
\lambda_{5 \alpha} \\
\delta \lambda_{4 \alpha} & =  0 = 2 \dot{\omega}_\alpha +\chi_\alpha
+ \epsab \omega_\beta \lambda_6
\end{array}
\right.
\eeq
which may be solved to obtain
\beq
\Rightarrow \quad \quad  \left\{  \begin{array}{rl}
\sigma & = - \dot{\xi} \\
h & =   \half \omega_\alpha \lambda_{5 \alpha}
- \half \ddot{\xi} \\
\chi_\alpha &= - 2 \dot{\omega}_\alpha - \epsab
\omega_\beta \lambda_6  .
\end{array}
\right.
\eeq
Notice there is a field-dependence in the parameters in the case of
$N=2.$  Plugging these into the infinitesimal transformations gives
\beqar
\delta x &=& \xi \dot{x} - \half \dot{\xi} x + \omega_\alpha
\psi_\alpha   \label{eq:inftran1} \\
\delta \psi_\alpha &=& \xi \dot{\psi}_\alpha  + \omega_\alpha \dot{x}
- \dot{\omega}_\alpha x + \epsab \omega_\beta  \left( F - \half
\lambda_6 x \right) - \half R \epsab \psi_\beta \\
\delta \lambda_2 &=& 2 \dot{\xi} \lambda_2 + \xi \dot{\lambda}_2
- \half \dddot{\xi}  + \frac{3}{2} \dot{\omega}_\alpha \lambda_{5
\alpha} + \half \omega_\alpha \dot{\lambda}_{5 \alpha}
+ \half \epsab \omega_\beta \lambda_6 \lambda_{5 \alpha}
\label{eq:lambda2} \\
\delta \lambda_{5 \alpha} &=& \frac{3}{2} \dot{\xi} \lambda_{5 \alpha}
+ \xi \dot{\lambda}_{5 \alpha} - 2 \ddot{\omega}_\alpha
+ 2 \omega_\alpha \left( \lambda_2 + \frac{1}{4} \lambda_6^{\:2} \right)
- \epsab \omega_\beta \dot{\lambda}_6  \\ \nonumber
& & - 2 \epsab \dot{\omega}_\beta \lambda_6
 - \half \epsab R \lambda_{5 \beta} \\
\delta \lambda_6 &=& \dot{\xi} \lambda_6 + \xi \dot{\lambda}_6
+ \dot{R} + \epsab \omega_\alpha \lambda_{5 \beta}  \label{eq:lambda6} \\
\delta F &=& \half \dot{\xi} F + \xi \dot{F}
-\epsab \omega_\alpha \dot{\psi}_\beta + \half \epsab \omega_\alpha
\lambda_{5 \beta} x + \half \omega_\alpha \psi_\alpha \lambda_6 .
\label{eq:inftran2}
\eeqar
These infinitesimal transformations have previously been obtained
by Siegel~\cite{siegel}.

\noindent
The case of $N=1$ is straightforward (here the auxiliary field
$F$ is not present, nor is the
gauge field $\lambda_6$ and its associated parameter $R$).
The bosonic and fermionic stress tensor correspond to $\lambda_2$
and $\lambda_5$ and there are matter fields $x$ and $\psi$ of weight
-1/2 and 0.
On the other hand, the infinitesimal algebra appears to be nonlinear
for $N=2$ !
The field dependence of the gauge parameters will generally introduce
terms quadratic in the gauge and matter fields.
It is easy to dismiss this illusion.
Firstly, we expect the matter be arranged in a supermultiplet.
This indicates that the auxiliary field is not $F,$ but is
\beq
\hat F  \equiv F - \half \lambda_6 x .
\eeq
Furthermore the terms
\beqarn
\delta \lambda_2 &\supset& \half \epsab \omega_\beta \lambda_{5 \alpha}
\lambda_6  \\
\delta \lambda_6 &\supset& - \half \epsab \omega_\beta \lambda_{5
\alpha}
\eeqarn
suggest that we take the combination $\lambda_2 + \frac{1}{4}
\lambda_6^{\:2} $ to be the spin two stress tensor for $N=2.$
With these identifications, the expressions above become linear.
Note that if the $F^2$ term in the action is rewritten in terms of
the primary field $\hat F,$ the shift in the stress tensor becomes
apparent
$$ \lambda_2 x^2 +  F^2 =
 \underbrace{ ( \lambda_2 + \frac{1}{4} \lambda_6^{\:2} ) }  x^2 +
 \hat{F}^2  + \hat F \lambda_6 x  .
$$

\section{Finite Superconformal Transformations}
Let us proceed with partial gauge fixing at the level of the finite Osp($N|2$)
transformations.  The finite transformations can be obtained by
exponentiation.  A recent discussion on finite gauge
transformations may be found in~\cite{gracia}.
The symmetries of the action in finite form are
\begin{itemize}

\item Reparametrizations
\beqarn
x^\prime(t)  &=& x(f(t))  \\
\psi^\prime_\alpha(t) &=& \psi_\alpha(f(t)) \\
F^\prime(t) &=& F(f(t))  \\
\lambda^\prime_A(t) &=& \dot{f}(t) \lambda_A(f(t)) \quad\quad
\forall \: A
\eeqarn

\item Scale Transformations
$$
\begin{array}{l}
x^\prime =  e^{\sigma /2}  \; x   \\
\psi_\alpha^\prime =   \psi_\alpha \\
F^\prime =  e^{- \sigma /2} \; F
\end{array} \quad \quad \quad
\begin{array}{l}
\lambda_1^\prime =  e^{\sigma} \; \lambda_1   \\
\lambda_2^\prime =  e^{- \sigma}  \; \lambda_2 \\
\lambda_3^\prime =  \lambda_3 + \dot{\sigma} \\
\lambda_{4 \alpha}^\prime =  e^{\sigma /2} \; \lambda_{4 \alpha} \\
\lambda_{5 \alpha}^\prime =  e^{-\sigma /2} \; \lambda_{5 \alpha} \\
\lambda_{6}^\prime =  \lambda_{6}
\end{array}
$$

\item Shift Transformations
$$
\begin{array}{l}
x^\prime = x  \\  \psi_\alpha^\prime = \psi_\alpha  \\
F^\prime = F
\end{array} \quad \quad \quad
\begin{array}{l}
\lambda_1^\prime = \lambda_1 \\
\lambda_2^\prime = \lambda_2 + \dot{h} + h \lambda_3 - h^2 \lambda_1 \\
\lambda_3^\prime = \lambda_3 - 2 h \lambda_1 \\
\lambda_{4 \alpha}^\prime = \lambda_{4 \alpha} \\
\lambda_{5 \alpha}^\prime = \lambda_{5 \alpha} - h \lambda_{4 \alpha} \\
\lambda_6^\prime =  \lambda_6
\end{array}
$$

\item Supersymmetries
\beqarn
x^\prime &=&  x + \omega_\alpha \psi_\alpha + \half \epsab
\omega_\alpha \omega_\beta F \\
\psi_\alpha^\prime &=& \psi_\alpha + \omega_\alpha \kappa
+ \epsab \omega F - \frac{1}{\lambda_1} \omega_\alpha \omega_\beta
S_{\psi_\beta} - \frac{F \lambda_{4 \gamma}}{4 \lambda_1} \left(
\epsilon_{\beta \gamma} \omega_\alpha \omega_\beta - \epsab
\omega_\gamma \omega_\beta \right) \\
F^\prime &=& F + \frac{1}{\lambda_1} \epsab \omega_\alpha
S_{\psi_\beta} - \frac{1}{2 \lambda_1} \omega_\alpha
\lambda_{4 \alpha} F + \frac{1}{2 \lambda_1} \epsab \omega_\alpha
\omega_\beta S_x \\
& & - \frac{1}{4 \lambda_1^{\:2}} \epsab \omega_\alpha
\omega_\beta \lambda_{4 \gamma} S_{\psi_\gamma}
+\frac{1}{4 \lambda_1^{\:2}} \left( \omega_\alpha \lambda_{4 \alpha}
\right)^2  \nonumber    \\
\lambda_1^\prime &=& \lambda_1 + \omega_\alpha \lambda_{4 \alpha}
+ \omega_\alpha \dot{\omega}_\alpha + \half \epsab \omega_\alpha
\omega_\beta \lambda_6 \\
\lambda_3^\prime &=& \lambda_3 + \omega_\alpha \lambda_{5 \alpha} \\
\lambda_{4 \alpha}^\prime &=& \lambda_{4 \alpha} + 2
\dot{\lambda}_\alpha - \omega_\alpha \lambda_3 + \epsab \omega_\beta
\lambda_6 - \omega_\alpha \omega_\beta \lambda_{5 \beta} \\
\lambda_{5 \alpha}^\prime &=& \lambda_{5 \alpha} + 2 \omega_\alpha
\lambda_2 \\
\lambda_6^\prime &=& \lambda_6 + \epsab \omega_\alpha \lambda_{5
\beta} + \epsab \omega_\alpha \omega_\beta \lambda_2
\eeqarn

\item Additional Fermionic Symmetries
\beqarn
x^\prime &=& x \\
\psi_\alpha^\prime &=&  \psi_\alpha + \half x \chi_\alpha \\
F^\prime &=& F \\
\lambda_1 &=& \lambda_1 \\
\lambda_2^\prime &=& \lambda_2 - \half \chi_\alpha \lambda_{5 \alpha}
- \frac{1}{8} \epsab \chi_\alpha \chi_\beta \lambda_6 - \frac{1}{4}
\chi_\alpha \dot{\chi}_\alpha \\
\lambda_3^\prime &=& \lambda_3 + \half \chi_\alpha \lambda_{4 \alpha} \\
\lambda_{4 \alpha}^\prime &=& \lambda_{4 \alpha} + \chi_\alpha
\lambda_1 \\
\lambda_{5 \alpha}^\prime &=& \lambda_{5 \alpha} + \half \chi_\alpha
\lambda_3 + \half \epsab \chi_\beta \lambda_6 + \dot{\chi}_\alpha
+ \frac{1}{4} \chi_\alpha \chi_\beta \lambda_{4 \beta} \\
\lambda_6^\prime &=& \lambda_6 + \half \epsab \chi_\beta \lambda_{4
\alpha} - \frac{1}{4} \epsab \chi_\alpha \chi_\beta \lambda_1
\eeqarn

\item o(2) Rotations
\beqarn
x^\prime &=& x \\
\psi_\alpha^\prime &=& ( \cos{\frac{R}{2}} \delta_{\alpha \beta}
+ \sin{\frac{R}{2}} \epsab ) \psi_\beta \\
F^\prime &=& F \\
\lambda_1^\prime &=& \lambda_1 \\
\lambda_2^\prime &=& \lambda_2 \\
\lambda_3^\prime &=& \lambda_3 \\
\lambda_{4 \alpha}^\prime &=& \left( \cos{\frac{R}{2}} \delta_{\alpha \beta}
+ \sin{\frac{R}{2}} \epsab \right) \lambda_{4 \beta} \\
\lambda_{5 \alpha}^\prime &=& \left( \cos{\frac{R}{2}} \delta_{\alpha \beta}
+ \sin{\frac{R}{2}} \epsab \right) \lambda_{5 \beta} \\
\lambda_6^\prime &=& \lambda_6 + \dot{R}
\eeqarn

\end{itemize}

\noindent
Perform successive finite transformations on each field ${\cal F}$
in the following (arbitrary but consistent order) manner:
\beq
{\cal F} \stackrel{\sigma}{\longrightarrow}
\Box \stackrel{h}{\longrightarrow}
\Box \stackrel{\chi_\alpha}{\longrightarrow}
\Box \stackrel{R}{\longrightarrow}
\Box \stackrel{\omega_\alpha}{\longrightarrow}
\Box \stackrel{f}{\longrightarrow}
\tilde{\cal F}.
\eeq

\noindent
For $N=1,$ we obtain the following finite Osp(1$|$2) transformed fields.
These are the active transformations of the fields.  The fields on the
left hand side with tildes are functions of $t$, while the fields and
parameters on the right hand side are functions of $f(t).$
\beqar
\tilde{x}(t) &=& e^{\sigma/2} \left( x + \omega \psi \right) \\
\tilde{\psi}(t) &=& \psi + \frac{\omega}{\lambda_1} \left(
\frac{\dot{x}}{\dot{f}} - \frac{\lambda_3 x}{2} - \frac{\lambda_4
\psi}{2} \right) +\frac{x}{2} \chi + \frac{ \omega \psi}{2} \chi \\
\tilde{\lambda_1}(t) &=& e^\sigma \dot{f} \left( \lambda_1 +
\omega \lambda_4 + \frac{\omega \dot \omega }{\dot f} \right) \\
\tilde{\lambda_2}(t) &=& e^{-\sigma} \dot f  \left( \lambda_2
+ \half \lambda_5 \chi + \omega \lambda_2 \chi
- \frac{\chi \dot{\chi} }{4 \dot f}  + \frac{\dot h}{\dot f} + h \lambda_3
+ h \omega \lambda_5
\right. \nonumber  \\ & & \left.
- \half h \lambda_4 \chi - \frac{h}{\dot f} \dot \omega \chi
+ \half h \lambda_3 \omega \chi - h^2 \lambda_1
- h^2 \omega \lambda_4  - h^2 \frac{\omega \dot \omega}{\dot f}
\right) \\
\tilde{\lambda_3}(t) &=& \dot{f} \left( \lambda_3 + \frac{\dot
\sigma}{\dot f} - 2 h \lambda_1 + \omega \lambda_5 - \half \lambda_4
\chi  - \frac{\dot{\omega}}{\dot f} \chi + \half \omega \lambda_3 \chi
-2 h \omega \lambda_4  \right.  \nonumber \\
& & \left. - 2 h \frac{\omega \dot \omega}{\dot f} \right) \\
\tilde{\lambda_4}(t) &=& e^{\sigma/2} \dot f \left( \lambda_4 +
2 \frac{\dot \omega}{\dot f} - \omega \lambda_3  + \lambda_1 \chi
+ \omega \lambda_4 \chi + \frac{\omega \dot \omega \chi}{\dot f}
\right) \\
\tilde{\lambda_5}(t) &=& e^{-\sigma/2} \dot f \left( \lambda_5 + 2 \omega
\lambda_2 + \frac{\dot \chi}{\dot f} + \half \lambda_3 \chi
+ \half \omega \lambda_5 \chi - h \lambda_4 - 2 h \frac{\dot
\omega}{\dot f}
\right.   \nonumber \\   & & \left.
+ h \omega \lambda_3 - h \lambda_1 \chi - h \omega
\lambda_4 \chi - h \frac{ \omega \dot \omega}{\dot f} \chi \right) .
\eeqar

\noindent
The gauge fixing conditions give the relations
\beq
\left\{ \begin{array}{cl}
\sigma &=  -\ln{\dot f} - \frac{\omega \dot \omega}{\dot f} \\
\chi &= - \frac{2 \dot \omega}{\dot f} \\
h &= \half \omega \lambda_5 + \omega \dot \omega \frac{\ddot
f}{\dot{f}^3} - \omega \ddot \omega \frac{1}{2 \dot{f}^2} -
\frac{\ddot f}{2 \dot{f}^2}
\end{array}
\right.
\eeq

\noindent
The resulting finite transformations of the matter fields $x$ and
$\psi$ and the gauge fields $T_B \equiv \lambda_2$ and
$T_F \equiv \half \lambda_5$ under the
residual symmetry are
\beqar
\tilde{x}(t) &=& \dot{f}^{-\half} \left( x + \omega \psi - \half
\frac{\omega \dot\omega}{\dot f} x \right) \\
\tilde{\psi}(t) &=& \psi + \frac{\omega}{\dot f}  \dot x
- \frac{\dot \omega}{\dot f}  x  + \frac{\omega \dot \omega}{\dot f}
\psi \\
\tilde{T}_B(t) &=& \dot{f}^2 \left[ T_B -\frac{\omega \dot \omega}{\dot f}
T_B + \frac{\omega}{\dot f} \dot{T}_F + 3 \frac{\dot \omega}{\dot f}
T_F \right]
-\half \left( \frac{\dddot f}{\dot f} - \frac{3 \ddot{f}^2}{2
\dot{f}^2} \right)
\\ \nonumber & &  + \half \omega \dot \omega
\frac{\dddot f}{\dot{f}^2} - \frac{\omega \dddot \omega}{2 \dot f}
- \frac{ 3 \ddot{f}^2}{2 \dot{f}^3} \omega \dot \omega
+ \frac{3 \ddot{f}}{2 \dot{f}^2} \omega \ddot \omega - \frac{3 \dot
\omega \ddot \omega }{2 \dot f} \\
\tilde{T}_F(t) &=& \dot{f}^{\frac{3}{2}} \left[ T_F + \omega T_B +
\frac{3 \omega \dot \omega }{2 \dot f}  T_F \right]
-\frac{1}{\sqrt{\dot f}} \left( \ddot \omega - \dot \omega \frac{\ddot
f}{\dot f} +  \frac{\omega \dot \omega \ddot \omega}{2 \dot f}
\right) .
\eeqar
These equations give the most general $N=1$ superconformal finite
transformations of the matter and gauge fields.
This procedure may unambiguously be carried out for any $N$
using the steps outlined here for $N=1.$
We will not write out the expressions for $N=2$ here, but
instead make  use of the superfield formulation to give
an alternate derivation of the $N=2$ superschwarzian.

\section{Superfield Formulation for $N=1,2$ and
Alternate Derivation of the Superschwarzian}

\subsection{$N=1$ Superfield Formulation}
Examination of the results in the previous section for $N=1$ show that
the matter fields and the gauge fields form supermultiplets
\beqar
\phi(Z) &=&  x(Z)  + \theta \psi(Z)  \nonumber \\
T(Z) &=& T_F(Z) + \theta T_B(Z)
\eeqar
\noindent
where $Z= (z,\theta)$ denotes the supercoordinates~\cite{friedan}.
Here we make a notational switch from $t$ in the 1-dimensional
particle model to $z$ to suggest a chiral sector of a two-dimensional
conformal theory.
The component transformations given in the previous section
are equivalent to the superfield transformations
\beqar
\phi(Z) &=&  \phi^\prime(Z^\prime) \left( D \theta^\prime \right)^{2h}
\quad \hbox{where } h= -\half  \nonumber \\
T(Z) &=&  T^\prime(Z^\prime) \left( D \theta^\prime \right)^3 +
\frac{\hat{c}}{4} \;
S(Z,Z^\prime)  \quad \hbox{where } \hat{c} = -4
\eeqar
where the superconformal derivative is
$$
D = \frac{\partial \:}{\partial \theta}
+ \theta \frac{\partial\:}{\partial z}
$$
and $S(Z,Z^\prime)$ is the superfield superschwarzian, which we will
address momentarily.
Superconformal transformations are those which induce a homogeneous
transformation law for the superconformal derivative.
{}From $D = (D \theta^\prime) D^\prime + (D z^\prime - \theta^\prime
D \theta^\prime ) \partial_z^\prime,$  we deduce that the
superconformal transformations are
\beqarn
z^\prime(z,\theta) &=& f(z) + \theta \omega(z) \sqrt{\partial_z f(z)} \\
\theta^\prime(z, \theta) &=& \omega(z) + \theta \sqrt{ \partial_z f(z)
+ \omega(z) \partial_z \omega(z) }.
\eeqarn

Using the superfield formulation, we may give an alternate
derivation of the superschwarzian in terms of superfields
rather than components, which is far simpler than any other method.
The equations of motion for the matter fields in the partially
gauge fixed pseudoparticle model are
\beqar
S_x &=& - \partial_z^{\:2} x + T_B x - T_F \psi  \nonumber \\
S_\psi &=&  - \partial_z \psi + T_F x .
\eeqar
These may be expressed as a superfield equation
\beq
\left( D^3 - T(Z) \right) \phi(Z) = 0.
\eeq
The schwarzian may be derived by demanding that this equation
maintain covariance under superconformal transformations.
{}From the infinitesimal transformation equation~\ref{eq:lambda2}
for $\lambda_2,$ we see that $c=-6$ or in superfield notation
$\hat c = \frac{2}{3} c = -4.$
The knowledge of the central charge allows us to normalize the
super schwarzian.
We write the anomalous term in the superfield transformation of the
super stress tensor as an apriori unknown function ${\cal S}(Z,Z^\prime).$
Writing the passive transformations
\beqarn
\phi^\prime(Z^\prime) &=& (D \theta^\prime) \phi(Z)  \\
D^\prime &=& (D \theta^\prime)^{-1} D \\
T^\prime(Z^\prime) &=& (D \theta^\prime)^{-3} (T(Z) + {\cal S}(Z,Z^\prime) )
\eeqarn
and plugging the expressions into the equation of motion, we obtain
\beq
\left( (D^\prime)^3 - T^\prime(Z^\prime) \right) \phi^\prime(Z^\prime)
= (D \theta^\prime)^{-2} \left( D^3 - T(Z) \right) \phi(Z)
\eeq
where
$$
{\cal S}(Z,Z^\prime) = \frac{D^4 \theta^\prime}{D \theta^\prime} -
2 \frac{D^3 \theta^\prime}{D \theta^\prime} \frac{D^2 \theta^\prime}{D
\theta^\prime}
$$
in agreement with Friedan~\cite{friedan}.

\subsection{$N=2$ Superfield Formulation}

We may repeat the above discussion for $N=2.$
Using complex notation  and the identifications
\beqar
T_B &=& \lambda_2 + \frac{1}{4} \lambda_6^{\:2}  \\
G^+ &=& \half \lambda_{5+} \\
G^- &=& - \half \lambda_{5-} \\
H &=& \half i \lambda_6
\eeqar
the superfields are, recalling that the correct matter
auxiliary field is  $\hat F = F - \half \lambda_6 x$
\beqar
\phi(Z) &=& x(Z) + \theta^+ \psi^-(Z) + \theta^- \psi^+(Z)
+ \theta^+ \theta^- \hat F(Z)  \nonumber \\
T(Z) &=& H(Z) + \theta^+ G^-(Z) + \theta^- G^+(Z) - \theta^+ \theta^- T_B(Z) ,
\eeqar
where $Z=(z,\theta^+,\theta^-)$ denotes the supercoordinates.
The equations of motion in components are
\beqar
S_x &=& - \partial_z^{\:2}  x + \left( T_B
+ \frac{1}{4} \lambda_6^{\:2} \right) x - G^+ \psi^- + G^- \psi^+ \\
S_{\psi_\pm} &=&  -\partial_z \psi_\pm + x G^\pm \pm H \psi^\pm \\
S_{\hat F}  &=& \hat F + \half \lambda_6 x
\eeqar
(the last equation is just equivalent to the equation $F=0$).
Using the superconformal derivatives
$$
D^\pm = \frac{\partial \:}{\partial \theta_\mp} + \theta^\pm
\partial_z,
$$
the pseudoparticle equations of motion can be written as the
superfield equation
\beq
\left[ \half \left( D^+ D^- - D^- D^+ \right) + T(Z) \right] \phi(Z) = 0.
\eeq
The $N=2$ super derivative transforms as
$$
D^\pm = (D^\pm \tilde{\theta}^-) \tilde{D}^+
      + (D^\pm \tilde{\theta}^+) \tilde{D}^-
+ \left( D^\pm \tilde{z} -  \tilde{\theta}^+ D^\pm \tilde{\theta}^-
                 -  \tilde{\theta}^- D^\pm \tilde{\theta}^+ \right)
\partial_{\tilde{z}} ,
$$
so that superconformal transformations must satisfy
$ D^\pm \tilde{z} -  \tilde{\theta}^+ D^\pm \tilde{\theta}^-
                 -  \tilde{\theta}^- D^\pm \tilde{\theta}^+ =0
$
and either $D^\pm \tilde{\theta}^\mp =0 $ or $D^\pm \tilde{\theta}^\pm =0.$
Choosing the superconformal condition
$$
D^\pm \tilde{\theta}^\pm = 0
$$
leads to the transformations in terms of $f(z),$ $\omega^\pm(z)$ and $R(z)$
\beqarn
\tilde{z}(z,\theta^+,\theta^-) &=& f
+ \theta^+ \omega^- e^{-\frac{i R}{2} }
\sqrt{\partial_z f + \omega^+ \partial_z \omega^-
                   + \omega^- \partial_z \omega^+}   \\
& & + \theta^- \omega^+ e^{\:\frac{i R}{2}}
\sqrt{\partial_z f + \omega^+ \partial_z \omega^-
                   + \omega^- \partial_z \omega^+}
 + \theta^+ \theta^- \partial_z ( \omega^+ \omega^- )  \\
\tilde{\theta}^+(z,\theta^+,\theta^-)
&=& \omega^+ + \theta^+ \sqrt{\partial_z f + \omega^+ \partial_z \omega^-
+ \omega^- \partial_z \omega^+} e^{-i\frac{R}{2}}  + \theta^+ \theta^-
\partial_z \omega^+ \\
\tilde{\theta}^-(z,\theta^+,\theta^-)
&=& \omega^- + \theta^- \sqrt{\partial_z f +
\omega^+ \partial_z \omega^- + \omega^- \partial_z \omega^+}
e^{i\frac{R}{2}}  + \theta^- \theta^+ \partial_z \omega^- .  \nonumber
\eeqarn
{}From the infinitesimal gauge fixing, the conformal weights
of $\phi$ and $T$ are determined, as well as the central charge
of the theory $\hat c = 1$.  We write passive transformation of the
super derivatives and fields as
\beqarn
\tilde{D}^+ &=& (D^+ \tilde{\theta}^-)^{-1} D^+    \\
\tilde{D}^- &=& (D^- \tilde{\theta}^+)^{-1} D^-    \\
\tilde{\phi}(\tilde Z) &=& (D^+ \tilde{\theta}^-)^{\half}
                 (D^- \tilde{\theta}^+)^{\half} \phi(Z)   \\
\tilde{T}(\tilde Z) &=&  (D^+ \tilde{\theta}^-)^{-1}
(D^-\tilde{\theta}^+)^{-1} \left[ T(Z) - {\cal S}(Z,\tilde Z) \right]
\eeqarn
where ${\cal S}(Z,\tilde{Z})$ is apriori an unknown function.
By demanding covariance of the superfield equation of motion
\beqar
\left[ \half \left( \tilde{D}^+ \tilde{D}^-  \right. \right. & - &
\left. \left.  \tilde{D}^-
\tilde{D}^+ \right)  + \tilde{T}(\tilde{Z}) \right]
\tilde{\phi}(\tilde Z) \\ \nonumber
&=& (D^+ \tilde{\theta}^-)^{-\half} (D^- \tilde{\theta}^+)^{- \half}
\left[ \half \left( D^+ D^- - D^- D^+ \right) + T(Z)  \right] \phi(Z)
\eeqar
one obtains the $N=2$ superschwarzian
\beq
{\cal S} = \frac{\partial_z D^- \tilde{\theta}^+ }{ \: D^-
\tilde{\theta}^+ }
       - \frac{\partial_z D^+ \tilde{\theta}^- }{\: D^+
\tilde{\theta}^- }
   + 2 \frac{\partial_z \tilde{\theta}^+ }{ D^- \tilde{\theta}^+ }
       \frac{\partial_z \tilde{\theta}^- }{ D^+ \tilde{\theta}^- }
\eeq
in agreement with the result of Cohn~\cite{cohn}.

\section{Complete gauge fixing of the pseudoparticle model}
Finally we perform the complete  gauge fixing of the model,
i.e.  in addition to setting $\lambda_1 = 1 $ and $\lambda_3 =
\lambda_{4 \alpha} = 0,$ we further fix $\lambda_2=0,$
$\lambda_{5 \alpha} =0$ and $\lambda_6 =0.$
Upon complete gauge fixing, the lagrangian reduces to
\beq
S = \int \: dt \left[ \half \dot{x}^2 +
\frac{1}{2} \dot{\psi}_\alpha \psi_\alpha  +  \half F^2 \right]  .
\eeq
Consequently the resulting pseudoparticle model will be governed
by the equations of motion
\beq
\ddot x = 0    \quad \quad  \dot \psi_\alpha = 0  \quad \quad   F=0
\eeq
which are invariant under the super M\"{o}bius transformations.
These are the transformations which are compatible with
putting the gauge field variations equal to zero, i.e. those
transformations for which the superschwarzian vanishes.
At the infinitesimal level for $N=1,$ we have
$\delta \lambda_2 = - \half \dddot{\xi} $ and $\delta \lambda_5  =
- 2 \ddot{\omega}$ so that $\xi(t) = A + t B + t^2 C$ and
$\omega(t) = \alpha + t \beta ,$ where $A,B,C$ and $\alpha,\beta$
are bosonic and fermionic constants.  At finite level, we have
\beqar
f(t) &=& \frac{a t + b}{c t + d}
\quad \quad \hbox{where} \;\; ad-bc=1  \nonumber \\
\omega(t) &=& \frac{\gamma t + \delta }{c t + d}
= (\gamma t + \delta) \sqrt{\dot f}
\eeqar
or in terms of superfields
(using $t^\prime(t,\theta) = f(t) + \theta \omega(t) \sqrt{\dot f(t)}$
and $\theta^\prime(t,\theta) = \omega(t) + \theta \sqrt{{\dot f}(t)
+ \omega(t) {\dot \omega}(t) }$)
\beq
t^\prime(t,\theta) =\frac{a t + b}{c t + d}  +
\theta \frac{\gamma t + \delta }{(c t + d)^2} \quad \quad \quad \quad
\theta^\prime(t,\theta) = \frac{\gamma t + \delta}{c t + d}
+ \theta \frac{1 + \delta \gamma /2 }{c t + d} .
\eeq
Similarly, for $N=2,$ the infinitesimal transformations are
$\delta \lambda_2 = - \half \dddot{\xi},$
$\delta \lambda_{5 \alpha}   = - 2 \ddot{\omega}_\alpha$ and
$\delta \lambda_6 = \dot{R}$ so that $\xi(t) = A + t B + t^2 C,$
$\omega^\pm(t) = \alpha^\pm + t \beta^\pm$  and $R(t)=R_0$ is a constant.
At the finite level, we find
\beqar
f(t) &=& \frac{a t + b}{c t + d}
\quad \quad \hbox{where} \;\; ad-bc=1  \nonumber \\
\omega^\pm(t) &=& \frac{\gamma^\pm t + \delta^\pm }{c t + d}
= (\gamma^\pm t + \delta^\pm) \sqrt{\dot f} \nonumber \\
R(t) &=& R_0.
\eeqar
The superfield expressions for the $N=2$ super M\"{o}bius transformations
may be obtained by plugging the above into the expressions in the previous
section for general superconformal transformations.

\section{$N=2$ twisted and topological theories}
In this section, we comment on twisted and topological
theories in relation to the $N=2$ pseudoparticle model.
{}From the pseudoparticle model point of view, the twisting of the
stress-tensor is a significant deformation of the untwisted theory.
Before elaborating on this point, we return to the infinitesimal
gauge fixing of the $N=2$ theory to show the emergence of the twisted
and topological theories.
{}From the infinitesimal transformation equations~\ref{eq:lambda2}
and~\ref{eq:lambda6},  one may define a twisted stress tensor
\beq
T_B = \lambda_2 + \frac{1}{4} \lambda_6^{\:2} + i \alpha_0 \dot \lambda_6 ,
\eeq
with which one may obtain a closed and consistent twisted $N=2$ algebra.
It will also prove useful to allow a shift of the parameter $R$
\beq
R = i \varphi \dot \xi + r .
\eeq
The transformations for gauge fields and the matter fields are
\beqar
\delta T_B &=& 2 T_B \dot \xi + \dot T_B \xi  - ( \half + \alpha_0
\varphi )  \dddot \xi + i \alpha_0  \ddot r
( \frac{3}{2} - \alpha_0 ) \dot \omega_+ \lambda_{5-}
\\ \nonumber
& &  + ( \frac{3}{2} + \alpha_0 ) \dot \omega_- \lambda_{5+}
 + ( \frac{1}{2} - \alpha_0 ) \omega_+ \dot \lambda_{5-}
 + ( \frac{1}{2} + \alpha_0 ) \omega_- \dot \lambda_{5+} \\
\nonumber
& &  + \left( \half \dot r + i ( \alpha_0 + \frac{\varphi}{2} )
\ddot \xi \right) \lambda_6  \\
\delta \lambda_{5 \pm}  &=& \left( \frac{3}{2} \mp \frac{\varphi}{2}
\right) \lambda_{5 \pm} \dot \xi +   \dot \lambda_{5 \pm} \xi
\pm \frac{i}{2} r  \lambda_{5 \pm} + 2 \omega_\pm T_B -2 \ddot
\omega_\pm
\\ \nonumber & &
\pm i \left( 1 \mp 2 \alpha_0 \right) \omega_\pm \dot
\lambda_6   \pm 2 \; i \;  \dot \omega_\pm \lambda_6  \\
\delta \lambda_6 &=& \dot \lambda_6 \xi + \lambda_6 \dot \xi + \dot r
+ i \varphi \ddot \xi + i \left( \omega_+ \lambda_{5-} - \omega_-
\lambda_{5+} \right) \\
\delta x &=&  \xi \dot x - \half \dot \xi x + \omega_- \psi_+
+ \omega_+ \psi_- \\
\delta \psi_\pm &=& \xi \dot{\psi}_\pm \mp \half \varphi \dot \xi
\psi_\pm + \omega_\pm \dot x - \dot{\omega}_\pm x
\mp \; i \; \omega_\pm \hat F  \pm  \frac{i }{2} r  \psi_\pm  \\
\delta \hat F &=& \xi \dot{\hat F} + \half \dot \xi \hat F
- \frac{i \varphi}{2} \ddot \xi x - \frac{\dot r}{2} x
+ i ( \omega_- \dot{\psi}_+  - \omega_+ \dot{\psi}_- ) .
\eeqar
{}From the transformation law for $T_B,$ it is evident that
for ordinary diffeomorphisms one should take $\varphi = - 2 \alpha_0.$
For generic $\alpha_0,$ the matter fields $x,$ $\psi_\pm$
and the fermionic gauge fields $\lambda_{5 \pm}$ transform as
primary fields under diffeomorphisms with weights $-\half,$
$\pm \alpha_0$ and $\frac{3}{2} \pm \alpha_0$ respectively.
The gauge fields $T_B,$ $\lambda_6$ and the auxiliary field
are quasi-primary with weights 2, 1 and $\half.$
The classical central charge of this theory is $c = -6 + 24
\alpha_0^{\:2}.$

Of particular interest is the point $\alpha_0^{\:2} = \frac{1}{4}$
where the central charge vanishes.  This is the $N=2$ topological theory.
With $\alpha_0 = - \half, $ we find
\beqar
\delta T_B &=& 2 T_B \dot \xi + \dot T_B \xi
-i \; \ddot r \dot{\omega_+}  \lambda_{5-} +
\half \dot r \lambda_6  +  2 \dot \omega_+ \lambda_{5-}
+ \dot \omega_- \lambda_{5+}  + \omega_+ \dot \lambda_{5-}  \\
\delta \lambda_{5 +} &=&\lambda_{5+} \dot \xi + \dot \lambda_{5+}\xi +
\frac{i}{2} r  \lambda_{5+} + 2 \omega_+ T_B -2 \ddot \omega_+
+ 2 i \dot \omega_+ \lambda_6  \\
\delta \lambda_{5 -} &=& 2 \lambda_{5-} \dot \xi +\dot \lambda_{5-}\xi
-\frac{i}{2} r  \lambda_{5-} + 2 \omega_- T_B -2 \ddot \omega_-
- 2 i \dot \omega_- \lambda_6  + 2 i \omega_- \dot \lambda_6 \\
\delta \lambda_6 &=& \dot \lambda_6 \xi + \lambda_6 \dot \xi + \dot r
+ i  \ddot \xi + i \left( \omega_+ \lambda_{5-} - \omega_-
\lambda_{5+} \right) \\
\delta x &=&  \xi \dot x - \half \dot \xi x + \omega_- \psi_+
+ \omega_+ \psi_- \\
\delta \psi_\pm &=& \xi \dot{\psi}_\pm \mp \half \dot \xi
\psi_\pm + \omega_\pm \dot x - \dot{\omega}_\pm x
\mp \; i \; \omega_\pm \hat F  \pm  \frac{i }{2} r  \psi_\pm    \\
\delta \hat F &=& \xi \dot{\hat F} + \half \dot \xi \hat F
- \frac{i }{2} \ddot \xi x - \frac{\dot r}{2} x
+ i ( \omega_- \dot{\psi}_+  - \omega_+ \dot{\psi}_- ) .
\eeqar
Thus $T_B,$ $\lambda_{5+},$  $\lambda_{5-},$ $x,$ $\psi_+$ and
$\psi_-$ transform as primary fields of weights 2,1,2,$\half,$
$-\half$ and $\half$ respectively.  The U(1) current $\lambda_6$
and the auxiliary field are quasiprimary with weights 1 and
$\half$ respectively.

Finally, we  comment on these twisted models as related to the
original pseudoparticle models.  The key observation is that
adding the twisting term $\alpha_0 \dot \lambda_6 $ to the stress
tensor is changing the
pseudoparticle model in a dramatic way, since the added term implies
that $\lambda_6$ no longer acts simply as a Lagrange multiplier
field but now has some dynamical term in the action.

Furthermore, if we do a complete gauge fixing of the twisted
theories, then the action and matter equations of motion
remain as in the previous section, but the residual symmetries
of the model are different.
{}From the central terms in equations~\ref{eq:inftran1}-\ref{eq:inftran2},
we have
$$
\left\{ \begin{array}{cl}
0 &= (\half - 2 \alpha_0^{\:2})  \dddot \xi  \\
0 &= \omega_\alpha \\
0 &= \dot r  -2 i \: \alpha_0 \ddot \xi .
\end{array}
\right.
$$
For nonzero $\alpha_0 \neq -\half$, the residual infinitesimal
transformations are $\xi = A + B t + C t^2,$
$\omega_\alpha = \gamma + \delta t$ and  $r = 4 \; i \alpha_0 C t + r_0.$
For the topological theory where $\alpha_0 = -\half$, then
$\xi = A + B t,$ $\omega_\alpha = \gamma + \delta t$ and $r = r_0.$
Thus the pseudoparticle model allows one to see that the
$N=2$ topological theory is in some sense
disconnected from the non-topological $N=2$ theory.

\section{Flag bundle interpretation}
In this section we describe how one may interpret, via the particle
model or the zero curvature equation, the superconformal transformations
as deformations of flags in superjet bundles over super Riemann surfaces.
This discussion will follow the general line of thought of
Gerasimov, Levin and Marshakov~\cite{geralm}
where the case of ${\cal W}_3$ was investigated, and of
Bilal, Fock and Kogan~\cite{bilal}.
Let us now explicitly construct the flags in the superjet
bundle for the cases of $N=1$ and $N=2.$  We will see that the pseudoparticle
model can serve as a guideline for how to perform this construction.

\subsection{$N=1$ Super Flag}
Consider the $-2h$ power of the canonical super line bundle over a
super Riemann surface, where $-2h =1 $ is the factor
appearing in the matter superfield transformation law.
A section ${\bf f} = {\bf f}(Z)$ of this super bundle is just a $-2h=1$
super differential.
Under superconformal transformations, ${\bf f}$ will be transformed
into an expression containing
${\bf f}$, $D {\bf f}$ and $ D^2 {\bf f}$.
For this reason, we should consider the $N=1$ super 1-jet
bundle~\footnote{ The term $N=1$ super 1-jet is used since the highest
bosonic derivative is $\partial_z,$ or equivalently, this is the $N=1$
supersymmetrization of an ordinary 1-jet.}
(two bosonic dimensions and one Grassmann dimension) over the
super Riemann surface.
A prolongation of the section ${\bf f}$ into a section of the
super 1-jet bundle, using a canonical basis is just
\beq
\hat {\bf f} ={\bf f} e_0 + (D {\bf f} ) e_1 +( D^2 {\bf f}) e_2 .
\eeq
We may alternately choose to express the section ${\bf f}$
using a different basis, determined by the solutions of the
completely gauge fixed pseudoparticle superfield matter equation
$$
D^3 \phi(Z) = 0.
$$
Working with 1-super differentials the solutions are
\beq
1 dZ \quad \quad  z dZ  \quad \quad  \theta  dZ .
\eeq
In a general coordinate system, the basis elements are defined
by the solutions to
$$
\left( D^3 - T \right) \phi(Z) = 0
$$
and are thus
\beq
D \tilde{\theta}  dZ \quad \quad
\tilde{z} D \tilde{\theta}  dZ \quad \quad
\tilde{\theta} D \tilde{\theta}  dZ .
\eeq
One may reexpress this latter ``dynamical'' basis in terms of the
canonical basis.  Writing
\beqarn
\eta_0 &=& D \tilde{\theta}   \\
\eta_\half &=& \tilde{\theta} D \tilde{\theta}   \\
\eta_1 &=& \tilde{z} D \tilde{\theta}
\eeqarn
the dynamical basis in terms of the canonical basis is given by
\beq
\left( \begin{array}{c}
\hat{\eta}_0 \\  \hat{\eta}_\half \\  \hat{\eta}_1 \end{array} \right) =
\left(  \begin{array}{ccc}
\eta_0  & D \eta_0 & D^2 \eta_0 \\
\eta_\half  & D \eta_\half & D^2 \eta_\half  \\
\eta_1  & D \eta_1 & D^2 \eta_1  \end{array}  \right)
\left( \begin{array}{c}  e_0 \\  e_1 \\  e_2  \end{array}
\right) .
\eeq
Writing the section of the super 1-jet bundle in the two different bases
\beq
\hat {\bf f} = {\bf f} e_0 + (D {\bf f} ) e_1 +( D^2 {\bf f}) e_2
= {\bf f}_0 \eta_0 + {\bf f}_\half \eta_\half + {\bf f}_1 \eta_1
\eeq
and then reexpressing ${\bf f}, D {\bf f} , D^2 {\bf f}$ in terms
of the ${\bf f}_0 , {\bf f}_\half , {\bf f}_1 ,$ we find the following
\beqar
{\bf f} &=& D \tilde{\theta} {\cal X}  \label{eq:n1flag1}\\
D {\bf f} &=& D^2 \tilde{\theta} {\cal X} + \left( D \tilde{\theta}^2
- 2 \tilde\theta D^2 \tilde \theta \right){\cal Y} \\
D^2 {\bf f} &=& D^3 \tilde{\theta} {\cal X}
+ D \tilde \theta  D^2 \tilde{\theta} {\cal Y} +
\left( D \tilde{\theta}^3 - 2 \tilde{\theta} D \tilde{\theta}
D^2 \tilde{\theta} \right) {\cal Z}  \label{eq:n1flag2}
\eeqar
where
\beqar
\cal X &=&  {\bf f}_0 + {\bf f}_\half \tilde{\theta}+{\bf f}_1 \tilde{z} \\
\cal Y &=&  {\bf f}_\half + {\bf f}_1 \tilde{\theta} \\
\cal Z &=&  {\bf f}_1 .
\eeqar
Equations~\ref{eq:n1flag1}-\ref{eq:n1flag2} describe a super flag
in the $N=1$ super 1-jet space
\beq
{\cal F}_{(1|0)} \quad \subset \quad {\cal F}_{(1|1)}
                 \quad \subset \quad {\cal F}_{(2|1)}
\eeq
where
\beqarn
{\cal F}_{(1|0)} &=& \left\{ \cal X \right\}  \\
{\cal F}_{(1|1)} &=& \left\{ \cal X, \cal Y \right\} \\
{\cal F}_{(2|1)} &=& \left\{ \cal X, \cal Y , \cal Z \right\} .
\eeqarn
Superconformal transformations generate deformations of the flag
\beq
\left( \begin{array}{ccc}
1  & - \tilde{\theta}  & \tilde{z}      \\
0  & 1               & \tilde{\theta} \\
0  & 0               & 1  \end{array} \right)
\left( \begin{array}{c}  {\bf f}_0 \\ {\bf f}_\half \\  {\bf f}_1
        \end{array}  \right) =
\left(  \begin{array}{l}  {\bf f}_0 + {\bf f}_\half \tilde{\theta}
+ \tilde{z} {\bf f}_1 \\
{\bf f}_\half  + \tilde{\theta} {\bf f}_1 \\
{\bf f}_1  \end{array} \right) .
\eeq

\subsection{$N=2$ Super Flag}
The particle model or zero curvature condition dictates that
one should consider a $-(h,q) = (\half,0)$ superdifferential ${\bf f},$
i.e. the matter superfield transforms as a $(-\half,0)$ superfield.
Under $N=2$ superconformal transformations, this  section ${\bf f}$
will be transformed into an expression containing
${\bf f}$, $D^\pm {\bf f}$ and $ \half \{ D^+, D^- \} {\bf f}$.
The $N=2$ super 1-jet bundle is two-bosonic two-fermionic dimensional
space over the super Riemann surface.
The prolongation of the section ${\bf f}$ into a section of the
$N=2$ super 1-jet bundle may be written in the canonical basis or the
dynamical basis.  The latter is defined by the matter equation of motion
in the complete gauge fixed case.  The solutions of
\beq
\half  \left( D^+ D^- - D^- D^+ \right)  \phi(Z) = 0
\eeq
are just 1, $\tilde{z},$  $\tilde{\theta}^+$ and $\tilde{\theta}^-.$
In an arbitrary coordinate system, the basis differentials are
\beqar
\eta_0 dZ &=& (D^+ \tilde{\theta}^-)^\half (D^-
\tilde{\theta}^+)^\half dZ \\
\eta_+ dZ &=& \tilde{\theta}^+ (D^+ \tilde{\theta}^-)^\half
(D^- \tilde{\theta}^+)^\half dZ  \\
\eta_- dZ &=& \tilde{\theta}^- (D^+ \tilde{\theta}^-)^\half
(D^- \tilde{\theta}^+)^\half dZ  \\
\eta_1 dZ &=& \tilde{z} (D^+ \tilde{\theta}^-)^\half (D^-
\tilde{\theta}^+)^\half dZ,
\eeqar
and as with $N=1,$ the dynamical basis is defined through the
prolongation of the $\eta_i$ via the canonical basis.
Writing the section ${\bf f}$ in terms of the two bases
\beqar
\hat {\bf f} &=& {\bf f} e_0 + (D^+ {\bf f} ) e_- + (D^- {\bf f} ) e_+
+ \half \{ D^+, D^- \} {\bf f}  e_1  \\ \nonumber
&=& {\bf f}_0 \hat \eta_0   + {\bf f}_+ \hat \eta_-
  + {\bf f}_- \hat \eta_+   + {\bf f}_1 \hat \eta_1
\eeqar
and then reexpressing ${\bf f},$ $D^\pm {\bf f},$
$\half \{ D^+,D^- \}  {\bf f}$ in terms
of the ${\bf f}_0,$ ${\bf f}_\pm,$ ${\bf f}_1 ,$ we find the following
\beqar
{\bf f} &=&  \eta_0  {\cal X} \\
D^\pm {\bf f} &=& D^\pm \eta_0 {\cal X}
+ \frac{\eta_0}{D^\mp \tilde{\theta}^\pm }  \left[
D^\pm \left( \tilde{\theta}^- D^\mp \tilde{\theta}^\pm \right) {\cal Y}^+
\right.   \\  & & \left.  +
D^\pm \left( \tilde{\theta}^+ D^\mp \tilde{\theta}^\pm \right) {\cal Y}^-
\right] \nonumber \\
\half \{ D^+, D^- \} {\bf f}  &=& \partial_z \eta_0 {\cal X}
+ \eta_0 \left( \left[ D^+,D^- \right] (\tilde{\theta}^+
\tilde{\theta}^-) - D^+ \tilde{\theta}^- D^- \tilde{\theta}^+ \right)
{\cal Z}   \nonumber \\
& & - \eta \partial_z \theta^- {\cal Y}^+
    - \eta \partial_z \theta^+ {\cal Y}^-
\eeqar
where
\beqar
{\cal X} &=&  {\bf f}_0 + {\bf f}_+ \tilde{\theta}^-
+ {\bf f}_- \tilde{\theta}^+  + {\bf f}_1 \tilde{z}
\label{eq:n2flag1} \\
{\cal Y}^\pm &=&  {\bf f}_\pm + {\bf f}_1 \tilde{\theta} \\
{\cal Z} &=&  {\bf f}_1 .
\label{eq:n2flag2}
\eeqar
Equations~\ref{eq:n2flag1}-\ref{eq:n2flag2} describe
a flag in the $N=2$ super 1-jet space
\beq
{\cal F}_{(1|0)} \quad \subset \quad {\cal F}_{(1|2)} \quad \subset \quad
{\cal F}_{(2|2)}
\eeq
where
\beqarn
{\cal F}_{(1|0)} &=& \left\{ \cal X \right\}  \\
{\cal F}_{(1|2)} &=& \left\{ \cal X, {\cal Y}^+, {\cal Y}^- \right\} \\
{\cal F}_{(2|2)} &=& \left\{ \cal X, {\cal Y}^+, {\cal Y}^- , \cal Z
\right\} .
\eeqarn
Superconformal transformations generate deformations of these flags
\beq
\left( \begin{array}{cccc}
1  & - \tilde{\theta}^- & -\tilde{\theta}^+  & \tilde{z}         \\
0  & 1                  & 0                  & \tilde{\theta}^+  \\
0  & 0                  & 1                  & \tilde{\theta}^-  \\
0  & 0                  & 0                  & 1  \end{array} \right)
\left( \begin{array}{c}  {\bf f}_0 \\ {\bf f}_+ \\ {\bf f}_- \\  {\bf f}_1
        \end{array}  \right) =
\left(  \begin{array}{l}  {\bf f}_0 + {\bf f}_+ \tilde{\theta}^-
+ {\bf f}_- \tilde{\theta}^+ + \tilde{z} {\bf f}_1 \\
{\bf f}_+   + \tilde{\theta}^+ {\bf f}_1 \\
{\bf f}_-   + \tilde{\theta}^- {\bf f}_1 \\
{\bf f}_1  \end{array} \right) .
\eeq

\subsection{General case}
We conjecture that for the case of arbitrary $N,$ the superconformal
transformations will induce deformations of a super flag in the
$N$-extended 1-jet of the associated super Riemann surface.
The flag should be
\beq
{\cal F}_{(1|0)} \quad \subset \quad {\cal F}_{(1|N)} \quad \subset \quad
{\cal F}_{(2|N)}
\eeq
where the subspaces of the flag are
\beqarn
{\cal F}_{(1|0)} &=& \left\{ {\cal X} \right\}  \\
{\cal F}_{(1|N)} &=& \left\{ {\cal X}, {\cal Y}^1,\ldots, {\cal Y}^N \right\}
\\
{\cal F}_{(2|2)} &=& \left\{ {\cal X}, {\cal Y}^1,\ldots, {\cal Y}^N ,
{\cal Z}     \right\}
\eeqarn
and the coordinates in the jet space are
\beqar
{\cal X} &=&  {\bf f}_0 + {\bf f}_\alpha \tilde{\theta}^\alpha
+ {\bf f}_z \tilde{z}  \\
{\cal Y}^\alpha &=&  {\bf f}_\alpha + {\bf f}_z \tilde{\theta} \\
{\cal Z} &=&  {\bf f}_z .
\eeqar

The superconformal transformations generate deformations of the
superflag
\beq
\left( \begin{array}{ccccc}
1  & - \tilde{\theta}^1 &  \cdots & -\tilde{\theta}^N  & \tilde{z}  \\
0  & 1 & \cdots & 0 & \tilde{\theta}^1  \\
0 & 0 \ddots & 0 & \vdots \\
0  & 0 & \cdots & 1 & \tilde{\theta}^2  \\
0  & 0  \cdots  & 0                  & 1  \end{array} \right)
\left( \begin{array}{c}  {\bf f}_0 \\ {\bf f}_1 \\ \vdots \\ {\bf f}_N\\
{\bf f}_z    \end{array}  \right) =
\left(  \begin{array}{l}  {\bf f}_0 + {\bf f}_\alpha
\tilde{\theta}^\alpha  + \tilde{z} {\bf f}_z \\
{\bf f}_1   + \tilde{\theta}^1 {\bf f}_z \\
\vdots \\
{\bf f}_N   + \tilde{\theta}^N {\bf f}_z \\
{\bf f}_z
  \end{array} \right) .
\eeq

\section{Conclusions and Outlook}
We have presented a one dimensional constrained  pseudoparticle
mechanical model which may be written as an Osp($N|2M$) phase space
gauge theory.  The partial gauge fixing of these models
yields theories which may be interpreted as chiral sectors of
two-dimensional theories of matter coupled to superconformal
${\cal W}$-gravity.  Partial gauge fixing of the
Osp($N|2$) pseudoparticle mechanical model results in
supergravity theories with SO($N$) invariant $N$-extended
superconformal symmetry of Bershadsky and Knizhnik.
Written as a phase space gauge theory, the pseudoparticle model
explains the success of the two-dimensional zero curvature approach to
finding extended conformal algebras.  In terms of the particle model,
the zero curvature method is
essentially equivalent to the fact that the compatibility of the
matter equation of motion and matter gauge transformation law yields
the gauge field transformation law, even after partial gauge fixing.
The finite transformations of the matter and non-gauge fixed gauge
fields may be obtained by integrating the osp($N|2M$) transformations
after redefinitions of gauge parameters has been performed to put
transformations into ``standard'' form,
transforming the matter and gauge fields by successive Osp($N|2M$)
transformations, and finally performing the partial gauge fixing at
finite level.
We have carried this procedure explicitly for the cases of Osp(2$|$1)
and Osp(2$|$2), thus giving a new derivation of the $N=1$ and $N=2$
superschwarzian derivatives.
An alternate derivation of the superschwarzian is given by writing
the matter equation of motion as a superfield and demanding covariance
of the equation under superconformal transformations.
The component version of this derivation could be useful
in the case of ${\cal W}$-algebras if there is some sort of
``${\cal W}$-field'' structure analogous to superfield structure.
With regard to the matter content of these theories,
if one wishes to have a conformal theory where the stress tensor
transforms as a quasi-primary field, then the conformal weight and spin
of the matter fields occurring in these
supergravity theories is pinned down due to the rigidity of the
compatibility equation.   Thus, there appears to be an obstruction in
coupling matter with arbitrary conformal weight and spin to the
untwisted pseudoparticle model as it has been presented.
To arrive at twisted and topological theories, a dynamically
significant deformation of the original pseudoparticle model must be
made, i.e. some Lagrange multiplier gauge fields must be given dynamics.
Unlike the untwisted theory, in the twisted theories one has the
freedom to couple matter of arbitrary conformal weight to
supergravity, due to the appearance of the arbitrary twisting
parameter $\alpha_0.$

The pseudoparticle model facilitates an
the interpretation of the SO($N$) invariant $N$-extended superconformal
transformations as deformations of flags in the $N$-supersymmetrized
1-jet bundles over super Riemann surfaces.
The matter conformal weight and charge dictates what power of the
canonical super line bundle one should use to begin the construction.
The weight and charge are constrained by the requirement that
the gauge fields transform as quasi-tensors.  The matter equation
of motion then defines a dynamical basis in the $N$-supersymmetrized 1-jet
which allows one to write down the flag.
Superconformal transformations change the way spaces are embedded
in higher dimensional spaces in the flag.

This procedure that we have given for finding the finite transformations
should extend to $N \geq 3$ to obtain the finite
transformations of the nonlinear SO($N$)
$N$-extended superconformal theories.   There is no real modification
necessary in going to $N \geq 3,$ only the addition  of more auxiliary
fields to close the algebra.
More generally, this procedure should allow one to obtain the finite
superconformally extended  ${\cal W}$-transformations.
The difficulty in considering Osp($N|2M$) for $M > 1$ is that there
is not a well defined notion of ``standard''
${\cal W}$-transformations.  The non-supersymmetric case has been
investigated in~\cite{gomishkr}.

Finally, it is natural to suspect a relation of the pseudoparticle
model with integrable hierarchies of nonlinear equations such as the
generalized Korteweg-de-Vries hierarchies.
Another interesting question is whether the quantization of particle models
of this type make sense.
These questions are currently under investigation.

\section*{Acknowledgements}
\hspace{\parindent}%
J.G. is grateful to Prof. S. Weinberg for the warm hospitality at the
Theory Group of the University of Texas at Austin, and to the
Ministerio de Educacion y Cienca of Spain for a grant.

This research was supported in part by Robert A. Welch Foundation,
NSF Grant 9009850 and CICYT project no.  AEN89-0347.

\end{document}